\documentclass[10pt]{iopart} 
\usepackage{iopams}
\usepackage{bbm}
\usepackage{psfrag}
\usepackage[dvips]{graphicx}
 
\newcommand\one{{\mathbbm{1}}}

\newcommand\RR{{\mathbbm{R}}}
\newcommand\CC{{\mathbbm{C}}}

\newcommand{\arctanh}{\mathrm{arctanh}}

\begin{document} 

\title[Nonequivalence of ensembles in the anisotropic quantum Heisenberg model]{Nonequivalence of ensembles in the Curie-Weiss anisotropic quantum Heisenberg model} 

\author{Michael Kastner$^{1,2}$}
\address{$^1$ National Institute for Theoretical Physics (NITheP), Stellenbosch 7600, South Africa}
\address{$^2$ Institute of Theoretical Physics,  University of Stellenbosch, Stellenbosch 7600, South Africa}
\ead{kastner@sun.ac.za} 

\date{\today}
 
\begin{abstract}
The microcanonical entropy $s(e,m)$ as a function of the energy $e$ and the magnetization $m$ is computed analytically for the anisotropic quantum Heisenberg model with Curie-Weiss-type interactions. The result shows a number of interesting properties which are peculiar to long-range interacting systems, including nonequivalence of ensembles and partial equivalence. Furthermore, from the shape of the entropy it follows that the Curie-Weiss Heisenberg model is indistinguishable from the Curie-Weiss Ising model in canonical thermodynamics, although their microcanonical thermodynamics in general differs. The possibility of experimentally realizing quantum spin models with long-range interactions in a microcanonical setting by means of cold dipolar gases in optical lattices is discussed.
\end{abstract}

%\noindent{\it Keywords\/}: Solvable lattice models, Quantum phase transitions (Theory), Classical phase transitions (Theory), 

%---------------------
\section{Introduction}
%---------------------

Many general theorems in statistical physics, but also the majority of applications, are concerned with short-range interacting systems. Short-range can either mean an algebraic decay, $r^{-d+\alpha}$, of the interaction potential with distance $r$, spatial dimension $d$, and a negative constant $\alpha$, or a finite range of interaction. For proving theorems in statistical physics, short-range interactions are a very handy assumption, as they allow one to partition a large volume into smaller subvolumes while neglecting the surface effects of the subvolumes in the thermodynamic limit. This trick is at the core of the proofs of the existence of thermodynamic potentials and their convexity properties in the thermodynamic limit, and also of the equivalence of statistical ensembles like the microcanonical, the canonical, and the grandcanonical one.

But there are also physical factors which account to some extend for the focus on short-range interactions: statistical physics was originally invented by Boltzmann, Gibbs, and others for studying the behaviour of gases, liquids, and solids, and the fundamental interactions in such systems are typically of an electromagnetic kind. The presence of positive and negative charges causes screening effects which give rise to interactions that are effectively of short range. For almost all practical purposes, effective interactions of finite range approximate this situation excellently, and this accounts for the almost exclusive interest in short-range interactions in the community of condensed matter physicists. 

The situation is different in astrophysics where gravitational interactions are of relevance. Since masses are non-negative, screening effects do not occur, and gravitational interactions retain their long-range character. As a consequence, it was in the astrophysical context that peculiarities of the statistical physics of long-range interacting systems first attracted attention. Negative heat capacities were shown to exist in bounded self-gravitating gas spheres by Lynden-Bell and Wood \cite{LynWood68} in 1968, and Thirring explained this observation by relating it to the nonequivalence of microcanonical and canonical ensembles \cite{Thirring70}. Although equivalence of ensembles had been proven only for short-range interactions, it was tacitly assumed by most physicists to hold in general. Therefore it came as a surprise to many that equivalence does not necessarily hold for long-range systems, in particular in the presence of a discontinuous phase transition.

Within the condensed matter community, interest in long-range interacting systems arose somewhat later, and physicists with an interest in fundamental issues of statistical physics studied simple long-range toy models in order to explore the {\em terra incognita}\ beyond traditional short-range statistical physics. In order to study nonequivalence of ensembles, it became necessary to compute thermodynamic functions within the microcanonical ensembles. This is in general more difficult than computations in the canonical ensemble, but a number of exact microcanonical solutions of long-range systems have been reported (see \cite{CamDauxRuf09} for a review). The models used in these studies---most prominently among them the Hamiltonian Mean-Field Model \cite{AnRu95}---consist of classical spin variables, and in most cases the interactions are of Curie-Weiss-type, i.e.\ each particle interacts with every other at equal strength. Curie-Weiss-type interactions can be considered as a limiting case of extremely long-range interactions, decaying as $r^{-d+\alpha}$ in the limit $\alpha\nearrow d$.

For quantum systems with long-range interactions, only very few microcanonical calculations have been reported in the literature. A notable example is the paper by Pflug \cite{Pflug80} on gravitating fermions, where a negative specific heat is found for all negative values of the energy. In the present article, we consider a different class of quantum systems, namely long-range quantum spin systems. The main goal is to contribute towards the understanding of nonequivalence of ensembles in quantum spin systems. The model chosen for this study is the anisotropic quantum Heisenberg model with Curie-Weiss-type interactions as defined in \sref{sec:model}. This is a model of anisotropically interacting spin-$1/2$ degrees of freedom, and the Curie-Weiss-type interactions render it possible to find an analytic solution for the microcanonical entropy in the thermodynamic limit. It is one of the main purposes of this article to show how such an analytic calculation can actually be done, and the presentation in \sref{sec:micent} is therefore somewhat technical. In \sref{sec:Legendre}, the canonical Gibbs free energy is recovered from the microcanonical entropy by means of a Legendre-Fenchel transform. Although the canonical result was known previously, its recovery from the microcanonical entropy is instructive and useful for discussing nonequivalence of ensembles and related issues in the subsequent sections.

In the remaining sections, physical implications of the results are discussed. In \sref{sec:nonequivalence}, nonequivalence of ensembles is discussed. In particular, it is shown that microcanonical and canonical ensembles are nonequivalent for a certain range of anisotropy parameters of the Curie-Weiss quantum Heisenberg model. A related observation is that, for the same range of anisotropy parameters, the canonical Gibbs free energy of the Curie-Weiss quantum Heisenberg model is identical to that of a (classical or quantum) Curie-Weiss Ising model. These two models are therefore thermodynamically equivalent in the canonical ensemble, but not so in the microcanonical ensemble, as discussed in \sref{sec:thermoequiv}. The physical relevance of a microcanonical treatment of quantum spin systems is discussed in \sref{sec:latticegas}. It is argued that the conditions under which a microcanonical description is adequate can be realized in experiments with cold atoms in traps. Dipolar gases in optical lattices can be used to engineer a large variety of long-range interacting condensed matter Hamiltonians, including the anisotropic quantum Heisenberg model \cite{Micheli_etal06}, and the absence of a coupling to a heat bath renders the microcanonical ensemble appropriate for a statistical description. We conclude in \sref{sec:conclusions} with a summary and a few further remarks.

Some of the results reported in this article, and in particular a discussion of their relevance for experiments with cold gases in optical lattices, have been previously announced (without the details of the calculations) in a Letter \cite{Kastner10}.

%---------------------------------------------------------
\section{Curie-Weiss anisotropic quantum Heisenberg model}
\label{sec:model}
%---------------------------------------------------------

The model consists of $N$ spin-$1/2$ degrees of freedom, each of which is interacting with every other at equal strength (Curie-Weiss-type interactions). The corresponding Hilbert space $\mathcal{H}=(\CC^2)^{\otimes N}$ is the tensor product of $N$ copies of the spin-$1/2$ Hilbert space $\CC^2$, and the Hamiltonian operator is given by
\begin{equation}\label{eq:Hamiltonian}
H_h=-\frac{1}{2N}\sum_{k,l=1}^N \left(\lambda_1 \sigma_k^1 \sigma_l^1 + \lambda_2 \sigma_k^2 \sigma_l^2 + \lambda_3 \sigma_k^3 \sigma_l^3\right) - h\sum_{k=1}^N \sigma_k^3.
\end{equation}
The $\sigma_k^\alpha$ are operators on $\mathcal{H}$ and act like the $\alpha$-component of the Pauli spin-$1/2$ operator on the $k$th factor of the tensor product space $\mathcal{H}$, and like identity operators $\one_2$ on all the other factors,
\begin{equation}
\sigma_k^\alpha=\one_2\otimes\cdots\otimes\one_2\otimes\underbrace{\sigma^\alpha}_{\hspace{-5mm}\mbox{$k$th factor}\hspace{-5mm}}\otimes\one_2\otimes\cdots\otimes\one_2,\qquad\alpha\in\{1,2,3\}.
\end{equation}
The resulting commutation relation is
\begin{equation}\label{eq:commutator}
\bigl[\sigma_k^\alpha,\sigma_l^\beta\bigr]=2\rmi\, \delta_{k,l}\,\epsilon_{\alpha\beta\gamma}\sigma_k^\gamma,\qquad\alpha,\beta,\gamma\in\{1,2,3\},
\end{equation}
where $\delta$ denotes Kronecker's symbol and $\epsilon$ is the Levi-Civita symbol. The parameter $h$ in the Hamiltonian is the strength of an external magnetic field orientated along the $3$-axis, and the constants $\lambda_1$, $\lambda_2$, and $\lambda_3$ determine the coupling strengths in the various spatial directions and allow to adjust the degree of anisotropy. %Note that it is explicitely shown in \cite{Micheli_etal06} that long-range anisotropic quantum Heisenberg models are among the systems that can be engineered with cold polar molecules in optical lattices.
It is often convenient to introduce a collective spin operator $\bi{S}$ with components
\begin{equation}\label{eq:Sa}
S_\alpha=\frac{1}{2}\sum_{i=1}^N \sigma_i^\alpha,\qquad\alpha\in\{1,2,3\},
\end{equation}
which allows us to rewrite the Hamiltonian \eref{eq:Hamiltonian} in the form
\begin{equation}\label{eq:Hamiltoniancollective}
H_h=-\frac{2}{N}\left(\lambda_1 S_1^2 + \lambda_2 S_2^2 + \lambda_3 S_3^2\right) - 2hS_3.
\end{equation}

Certain choices for the coupling constants of the anisotropic Heisenberg model yield several important special cases, for example: (a) the isotropic Heisenberg model, $\lambda_1=\lambda_2=\lambda_3$; (b) the Ising model, $\lambda_1=0=\lambda_2$; (c) the isotropic Lipkin-Meshkov-Glick model, $\lambda_1=\lambda_2$ and $\lambda_3=0$. For these special cases, and more generally whenever $\lambda_1=\lambda_2$, the Hamiltonian \eref{eq:Hamiltoniancollective} can be expressed entirely in terms of $\bi{S}^2$ and $S_3$, i.e., the square and the 3-component of the collective spin. As a consequence, an angular momentum eigenbasis diagonalizes $H_h$ and $S_3$ simultaneously, and the model can be solved by rather elementary means, as shown in \sref{sec:special}

Here we consider the coupling constants $\lambda_1$, $\lambda_2$, and $\lambda_3$ to be nonnegative, but otherwise arbitrary. The exact expression for the canonical Gibbs free energy $g$ as a function of the inverse temperature $\beta=1/T$ \footnote{Here and in the following Boltzmann's constant is set to unity.} and the magnetic field $h$ is known for this model (and, in fact, for a larger class of systems) and is reported for example in \cite{PearceThompson75}. The model is found to display a transition from a ferromagnetic to a paramagnetic phase in the canonical ensemble.%, and the phase diagram is shown in \fref{fig:canPD}.

%-------------------------------
\section{Microcanonical entropy}
\label{sec:micent}
%-------------------------------
Owing to the long-range character of the interactions in the Hamiltonian \eref{eq:Hamiltonian}, microcanonical and canonical ensembles cannot be expected to yield equivalent results. The main purpose of the present article is therefore to complement the known canonical results with those obtained in the microcanonical ensemble.

In thermodynamics, the energy $e$ is the variable conjugate to the inverse temperature $\beta$, and the magnetization $m$ is conjugate to $-\beta h$. In the same way as $g(\beta,h)$ represents the fundamental quantity of the quantum Heisenberg model in the canonical ensemble, the microcanonical entropy $s(e,m)$ serves as a starting point for a microcanonical description in the thermodynamic limit. However, for a pair of variables $(e,m)$ corresponding to the pair of noncommuting operators $(H_0,M=\sum \sigma_i^3)$, it is not even well established how to define a quantum microcanonical entropy. Extending a suggestion of Truong \cite{Truong74} to interacting systems, the definition
\begin{equation}\label{eq:sNemdef}
s_N(e,m)\!=\!\frac{1}{N}\ln\sum_{\bar{e},\bar{m}}\Tr\left[P_{H_0}(\bar{e}) P_M(\bar{m}) \right] \delta_\Delta(\bar{e}-e)\delta_\Delta(\bar{m}-m)
\end{equation}
seems to be physically reasonable, but difficult to apply in practice. Here, $\bar{e}$ and $\bar{m}$ denote eigenvalues of the operators $H_0/N$ and $M/N$, respectively. $\delta_\Delta$ is the characteristic function of the interval $[-\Delta,0]$, i.e.,  $\delta_\Delta(x)=1$ if $x\in[-\Delta,0]$, and zero otherwise. The parameter $\Delta>0$ is chosen small but otherwise arbitrary. $P_{H_0}(\bar{e})$, $P_M(\bar{m})$ denote the eigenprojections of the operators $H_0$ and $M$ belonging to the eigenvalues $\bar{e}$ and $\bar{m}$, respectively.

Definition \eref{eq:sNemdef} can be motivated as follows: In classical physics, the density of states for a pair of variables is
\begin{equation}
\Omega_{\mathrm{classical}}(e,m)=\sum_x \delta_\Delta(\bar{e}(x)-e)\delta_\Delta(\bar{m}(x)-m),
\end{equation}
where the summation is over phase space (assumed here to be discrete for simplicity) and $\bar{e}$ and $\bar{m}$ are phase space functions. The interpretation of this classical expression is: Pick a state $x$ with energy $\bar{e}(x)$ in the interval $[e-\Delta,e]$. If furthermore  $\bar{m}(x)$ is in the interval $[m-\Delta,m]$, then add one to the sum, otherwise zero. Examples of such densities of states (or the corresponding entropies $s_{\mathrm{classical}}(e,m)=\ln\Omega_{\mathrm{classical}}(e,m)/N$) for classical spin systems can be found in \cite{Kastner02,HaKa06,Kastner09}.

For a quantum system, the definition has to be modified: Since the operators $H_0$ and $M$ do not commute in general, an eigenstate of $H_0$ is usually not an eigenstate of $M$. An eigenstate of $H_0$ with eigenvalue $\bar{e}$ in the interval $[e-\Delta,e]$ will therefore in general not have a well-defined value of $\bar{m}$. Instead, measuring the magnetization of this energy eigenstate can yield one out of several values of $\bar{m}$, each with a certain probability. It seems therefore reasonable to define the density of states [i.e.\ the argument of the logarithm in \eref{eq:sNemdef}] by adding up, for all energy eigenstates with eigenvalue $\bar{e}$ in the interval $[e-\Delta,e]$, the probabilities for finding a magnetization in the interval $[m-\Delta,m]$. For one-dimensional subspaces, this probability is given by the overlap $|\langle\bar{e}|\bar{m}\rangle|^2$, and for higher dimensional subspaces (i.e.\ degenerate eigenvalues) it can be expressed in terms of projection operators as in the trace in \eref{eq:sNemdef}. For the special case of commuting operators, these probabilities are either one or zero, and the density of states counts, in analogy to the classical case, the states that comply with a certain constraint.  

At least for finite systems, and in contrast to its classical counterpart, the entropy $s_N(e,m)$ defined in \eref{eq:sNemdef} does {\em not}\/ describe an ensemble of systems at fixed energy $e$ and magnetization $m$: For noncommuting operators $H_0/N$ and $M/N$, quantum mechanical uncertainty does not simultaneously allow fixed values for both quantities. %, the magnetization $m$ is not a conserved quantity under the quantum mechanical time evolution induced by $H_h$. 
It is, however, possible to give a probabilistic interpretation of the (suitably normalized) density of states,
\begin{equation}\label{eq:OmegaNemdef}
\Omega_N(e,m)\propto\sum_{\bar{e},\bar{m}}\Tr\left[P_{H_0}(\bar{e}) P_M(\bar{m}) \right] \delta_\Delta(\bar{e}-e)\delta_\Delta(\bar{m}-m),
\end{equation}
as the probability of measuring, with experimental resolution $\Delta$, a certain value of $m$ at a given value of the energy $e$.

In the thermodynamic limit $N\to\infty$, the familiar interpretation of $s_N(e,m)$ as describing an ensemble of systems at fixed energy $e$ and magnetization $m$ is recovered. One way to understand this property relies on an observation made by von Neumann in the early days of quantum mechanics (\cite{vonNeumann29}; see also \cite{Goldstein_etal} for a modern presentation of these ideas). Assume that an experimenter attempts to measure simultaneously the energy and the magnetization of a macroscopic system. Von Neumann argues that macroscopic observables do always commute, since nothing prevents the experimenter from reading out the display of his measurement devices simultaneously. However, the observables measured are not really energy and magnetization, but two commuting observables, approximating the observables of interest to a very good degree. De Roeck {\em et al.} \cite{DeRoeck_etal06} implemented this idea to define the microcanonical entropy of a quantum system in the thermodynamic limit, making use of the concept of concentrating sequences. In the same spirit, we would like to compute an entropy $s_N(\tilde{e},\tilde{m})$, where $\tilde{e}$ and $\tilde{m}$ correspond to commuting operators $\tilde{H}_0/N$ and $\tilde{M}/N$ which converge to $H_0/N$ and $M/N$ in the thermodynamic limit in a suitable way. However, an explicit construction of $\tilde{H}_0$ and $\tilde{M}$ is difficult, and it is more convenient to use, as in \eref{eq:sNemdef}, the observables $H_0$ and $M$, and the resulting entropy coincides with the desired one in the thermodynamic limit.
%can expect the thermodynamic limit of the entropy \eref{eq:sNemdef} to describe an ensemble of systems at fixed energy $e$ and magnetization $m$, since the relative strength of the quantum fluctuations becomes negligible. 

%----------------------------------------------------
\subsection{Special values of the coupling constants: $\lambda_1=\lambda_2$}
\label{sec:special}
%----------------------------------------------------
By defining collective spin ladder operators
\begin{equation}
S_-=\case{1}{2}\left(S_1-\rmi S_2\right),\qquad S_+=S_-^\dagger=\case{1}{2}\left(S_1+\rmi S_2\right),
\end{equation}
the Hamiltonian $H_h$ in \eref{eq:Hamiltoniancollective} can be written in the form
\begin{equation}
\fl H_h=-\frac{1}{N}\left[\left(\lambda_1+\lambda_2\right)\bi{S}^2-\left(\lambda_1+\lambda_2-2\lambda_3\right)S_3^2+2\left(\lambda_1-\lambda_2\right)\left(S_-^2+S_+^2\right)\right]-2hS_3.
\end{equation}
In the special case $\lambda_1=\lambda_2\equiv\lambda_\perp$, the term containing $S_-$ and $S_+$ vanishes, and the Hamiltonian can be expressed entirely in terms of the operators $\bi{S}^2$ and $S_3$,
\begin{equation}
\tilde{H}_h=-\frac{2}{N}\left[\lambda_\perp\bi{S}^2-\left(\lambda_\perp-\lambda_3\right)S_3^2\right]-2hS_3.
\end{equation}
The components $S_\alpha$ of the collective spin operator deserve their name, meaning that they obey the angular momentum algebra
\begin{equation}
\left[S_\alpha,S_\beta\right]=\rmi \epsilon_{\alpha\beta\gamma}S_\gamma,\qquad\alpha,\beta,\gamma\in\{1,2,3\},
\end{equation}
as is easily verified from relation \eref{eq:commutator} and definition \eref{eq:Sa}. Therefore, the well-known angular momentum eigenstates $|S,M\rangle$ are eigenstates of the Hamiltonian $\tilde{H}_h$ with eigenvalues
\begin{equation}
E_{SM,h}=-\frac{2}{N}\left[\lambda_\perp S(S+1)-\left(\lambda_\perp-\lambda_3\right)M^2\right]-2hM.
\end{equation}
From the rules for the addition of angular momenta it follows that $S$ can take on integer values from $0$ to $N/2$ where, for simplicity, we have assumed $N$ to be even. The Hilbert space can be decomposed into a direct sum of irreducible representations of $SO(3)$,
\begin{equation}
\mathcal{H}=(\CC^2)^{\otimes N}\cong\bigoplus_{S=0}^{N/2}d_S\mathcal{D}_S.
\end{equation}
Each representation $\mathcal{D}_S$ is a linear subspace of dimension $2S+1$ with basis $|S,M\rangle$, $M\in\{-S,\dots,+S\}$, and
\begin{equation}
d_S=\frac{N!\,(2S+1)}{(N/2+S+1)!(N/2-S)!},
\end{equation}
is the multiplicity of the representation $\mathcal{D}_S$ for $N$ even \cite{Scharf72}.

Now the trace in \eref{eq:sNemdef} can be written with respect to the $|S,M\rangle$-eigenbasis,
\begin{equation}
\Omega_N(e,m)=\sum_{S=0}^{N/2}d_S\sum_{M=-S}^{+S}\delta_\Delta\left(Ne-E_{SM,0}\right)\delta_\Delta\left(Nm-2M\right).
\end{equation}
In the limit of small $\Delta$ and large system size $N$, we can write
\begin{equation}
\eqalign{
s(e,m)=\lim_{N\to\infty}\frac{1}{N}\ln\Omega_N(e,m)\\
=\lim_{N\to\infty}\frac{1}{N}\ln\sum_{S=0}^{N/2}d_S\,\delta_\Delta\!\left(Ne+\frac{2\lambda_\perp S(S+1)}{N}-\frac{Nm^2\left(\lambda_\perp-\lambda_3\right)}{2}\right)\\
=\lim_{N\to\infty}\frac{1}{N}\ln\frac{N!\,2S}{(N/2+S)!(N/2-S)!}\Bigg|_{2S=N\sqrt{m^2(1-\lambda_3/\lambda_\perp)-2e/\lambda_\perp}}.
}
\end{equation}
Applying Stirling's formula, we obtain for the microcanonical entropy of the an\-iso\-trop\-ic quantum Heisenberg model in the thermodynamic limit the final expression
\begin{equation}\label{eq:sfinalspecial}
\fl s(e,m)=\ln2-\frac{1}{2}[1-f(e,m)]\ln[1-f(e,m)]-\frac{1}{2}[1+f(e,m)]\ln[1+f(e,m)]
\end{equation}
with 
\begin{equation}\label{eq:femspecial}
f(e,m)=\sqrt{m^2\left(1-\frac{\lambda_3}{\lambda_\perp}\right)-\frac{2e}{\lambda_\perp}},
\end{equation}
where $\lambda_\perp=\lambda_1=\lambda_2$. The domain of $s$ is given by
\begin{equation}\label{eq:domainspecial}
\fl\mathcal{D}=\left\{(e,m)\in\RR^2\,\big|\,2e+m^2\lambda_3<0\;\mbox{and}\;\lambda_\perp > m^2\left(\lambda_\perp-\lambda_3\right)-2e\right\}.
\end{equation}

%----------------------------------------------------
\subsection{General values of the coupling constants}
\label{sec:general}
%----------------------------------------------------
For positive, but otherwise arbitrary, values of the coupling constants $\lambda_\alpha$ in the Hamiltonian \eref{eq:Hamiltonian}, an evaluation of the expression \eref{eq:sNemdef} is difficult to achieve. Considering the thermodynamic limit of the entropy,
\begin{equation}\label{eq:sem2def}
s(e,m)=\lim_{N\to\infty} s_N(e,m),
\end{equation}
usually accounts for a simplification of the problem, but not enough so in this case. Here, in order to render the calculation feasible, we claim that the exact result of expression \eref{eq:sem2def} is recovered by formally computing, in a sense which will become clear in the following, the quantity
\begin{equation}\label{eq:sem3def}
s(e,m)=\lim_{N\to\infty}\frac{1}{N}\ln\Omega_N(e,m)
\end{equation}
with a density of states $\Omega_N$ that we write symbolically as
\begin{equation}\label{eq:Omegadef}
\Omega_N(e,m) = \Tr\left[\delta(e-H_0/N)\delta(m-M/N)\right].
\end{equation}
Note that the symbolic expressions make little mathematical sense and require some physically reasonable regularization, like the $\Delta$-regularization in \eref{eq:OmegaNemdef}. Here, instead, we will keep the unregularized deltas for the moment to render some formal manipulations more obvious. Expressing the deltas in their Fourier representation and performing the thermodynamic limit later on will cure the problem and lead to a well-defined mathematical expression for the microcanonical entropy $s(e,m)$ in the thermodynamic limit.

In view of the deltas, the integral remains unchanged upon insertion of the factors $\exp[a(Ne-H_0)]$ and $\exp[b(Nm-M)]$ in the trace,
\begin{equation}
\fl\Omega_N(e,m) = \Tr\left\{N^2\exp[a(Ne-H_0)]\delta(Ne-H_0)\exp[b(Nm-M)]\delta(Nm-M)\right\},
\end{equation}
where $a$ and $b$ are real parameters. Writing the deltas in Fourier representation, we obtain
\begin{equation}\label{eq:Omega1}
\eqalign{
\fl\Omega_N(e,m) = \frac{N^2}{4\pi^2}\int\rmd k \int\rmd \ell \Tr\biggl\{ \exp\left[(a+\rmi k)\left(Ne+\frac{2}{N}\left(\mathcal{S}_1^2+\mathcal{S}_2^2+\mathcal{S}_3^2\right)\right)\right]\\
\times\exp\left[(b+\rmi \ell)\left(Nm-2S_3\right)\right]\biggr\},
}
\end{equation}
where the anisotropic collective spin operators
\begin{equation}
\mathcal{S}_\alpha^2=\lambda_\alpha S_\alpha^2
\end{equation}
have been introduced to ease the notation. Unless specified otherwise, domains of integration always extend from $-\infty$ to $+\infty$.

\subsubsection{Decoupling the $N$-spin trace.}
To evaluate this expression, we want to rewrite the density of states \eref{eq:Omega1} in such a way that the $N$-spin trace decouples into a product of one-spin traces. To this purpose, we can adapt to our microcanonical setting a couple of tricks that have been used by Tindemans and Capel \cite{TinCa74} in a related canonical calculation. The first exponential in \eref{eq:Omega1} is an exponential of a sum of squares of the collective spin components $S_\alpha$, which---for a reason that will become clear soon---we would like to rewrite as a product of exponential operators. This can be done by using a straightforward generalization of the Lie-Trotter formula \cite{Trotter59}, yielding
\begin{equation}\label{eq:limexp1}
\eqalign{
\exp\left[(a+\rmi k)\left(Ne+\frac{2}{N}\left(\mathcal{S}_1^2+\mathcal{S}_2^2+\mathcal{S}_3^2\right)\right)\right]\\
= \lim_{n\to\infty}\left[\exp\left(\frac{(a+\rmi k)Ne}{n}\right)\prod_{\alpha=1}^3\exp\left(\frac{2(a+\rmi k)}{nN}\mathcal{S}_\alpha^2\right)\right]^n.
}
\end{equation}
Now the exponential operators are in a suitable form for applying the Hubbard-Stratonovich trick \cite{Stratonovich57,Hubbard59}. This is a fancy name for the identity
\begin{equation}
\rme^{cO^2} = \frac{1}{n\sqrt{\pi c}}\int\rmd x\, \rme^{-x^2/(cn^2)}\rme^{2xO/n},
\end{equation}
where $O$ is an operator and $c$ is a constant with $\Re(c)>0$. With this formula, we can express \eref{eq:limexp1} as
\begin{equation}\label{eq:limexp2}
\fl\lim_{n\to\infty}\left[\left(\frac{N}{2\pi n(a+\rmi k)}\right)^{3/2}\rme^{(a+\rmi k)Ne/n}\int\rmd^3 x \exp\left(-\frac{N\bi{x}\cdot\bi{x}}{2n(a+\rmi k)}\right)\prod_{\alpha=1}^3\rme^{2x_\alpha\mathcal{S}_\alpha/n}\right]^n,
\end{equation}
where $\bi{x}=(x_1,x_2,x_3)$. As a result, the operators $\mathcal{S}_\alpha$ now occur linearly in the exponent, which will be important for what follows. Next we replace the product (over $\alpha$) of three exponentials in \eref{eq:limexp2} by the single exponential
\begin{equation}\label{eq:replace1}
\exp\left(\frac{2}{n}\sum_{\alpha=1}^3 x_\alpha \mathcal{S}_\alpha\right).
\end{equation}
This term is in general not identical to the product in \eref{eq:limexp2}, but the correction---as given by the Baker-Campbell-Hausdorff formula---is of the order $1/n^2$ and should therefore in the limit $n\to\infty$ be negligible when compared to terms of the order $1/n$. A rigorous justification of the replacement \eref{eq:replace1} is given in Appendix A of \cite{TinCa74}.

Recalling that the operators $2\mathcal{S}_\alpha=\sqrt{\lambda_\alpha}\sum_{i=1}^N\sigma_i^\alpha$ are sums of one-particle-operators $\sigma_i^\alpha$, we can rewrite the exponential \eref{eq:replace1} as
\begin{equation}\label{eq:replace2}
\exp\left(\frac{1}{n}\sum_{\alpha=1}^3 x_\alpha \sqrt{\lambda_\alpha}\sum_{i=1}^N \sigma_i^\alpha\right) = \exp\left(-\frac{1}{n}\sum_{i=1}^N\xi(i,\bi{x})\right),
\end{equation}
where
\begin{equation}
\xi(i,\bi{x}) = -\sum_{\alpha=1}^3 x_\alpha \sqrt{\lambda_\alpha}\, \sigma_i^\alpha.
\end{equation}
Since the $\xi(i,\bi{x})$ are one-spin operators, acting non-trivially on only one of the factors of the tensor-product Hilbert space $(\CC^2)^{\otimes N}$, their commutators must vanish,
\begin{equation}
\left[\xi(i,\bi{x}),\xi(j,\bi{y})\right]=0\qquad\mbox{for $i\neq j$}.
\end{equation}
Hence, the exponential in \eref{eq:replace2} is equal to
\begin{equation}
\prod_{i=1}^N \rme^{-\xi(i,\bi{x})/n},
\end{equation}
and we obtain
\begin{equation}\label{eq:limexp3}
\fl\rme^{(a+\rmi k)Ne}\lim_{n\to\infty}\left[\left(\frac{N}{2\pi n(a+\rmi k)}\right)^{3/2}\int\rmd^3 x \exp\left(-\frac{N\bi{x}\cdot\bi{x}}{2n(a+\rmi k)}\right)\prod_{i=1}^N \rme^{-\xi(i,\bi{x})/n}\right]^n
\end{equation}
for the expression in \eref{eq:limexp2}. Now we can express the $n$th power of a three-dimensional integral in \eref{eq:limexp3} as a $3n$-dimensional integral,
\begin{equation}\label{eq:limexp4}
\eqalign{
\fl\rme^{(a+\rmi k)Ne}\lim_{n\to\infty}\left(\frac{N}{2\pi n(a+\rmi k)}\right)^{3n/2} \int\!\cdots\!\int\rmd^3 x^{(1)}\dots\rmd^3 x^{(n)}\\
\times\prod_{m=1}^n \exp\left(-\frac{N\bi{x}^{(m)}\cdot\bi{x}^{(m)}}{2n(a+\rmi k)}\right)\prod_{i=1}^N \rme^{-\xi\left(i,\bi{x}^{(m)}\right)/n}.
}
\end{equation}
Inserting this expression into the density of states \eref{eq:Omega1} and making use of the linearity of the trace, we can write
\begin{equation}\label{eq:Omega2}
\eqalign{
\fl\Omega_N(e,m) = \frac{N^2}{4\pi^2}\int_{a-\rmi\infty}^{a+\rmi\infty}\rmd s \,\rme^{Nes} \int_{b-\rmi\infty}^{b+\rmi\infty}\rmd t\, \rme^{Nmt} \lim_{n\to\infty}\Biggl\{\left(\frac{N}{2\pi ns}\right)^{3n/2}\\
\times \int\!\cdots\!\int\rmd^3 x^{(1)}\cdots\rmd^3 x^{(n)} \exp\left(-\frac{N}{2ns}\sum_{m=1}^n \bi{x}^{(m)}\cdot\bi{x}^{(m)}\right)\\
\times\Tr\left[\left(\prod_{m=1}^n \prod_{i=1}^N \rme^{-\xi\left(i,\bi{x}^{(m)}\right)/n}\right) \rme^{-2tS_3}\right]\Biggr\}.
}
\end{equation}
Here we have also substituted the integration variables $k$ and $\ell$ by $s=a+\rmi k$ and $t=b+\rmi\ell$. The trace in \eref{eq:Omega2} acts only on exponentials of one-particle operators, and this allows us to rewrite the trace $\Tr$ on the $N$-spin Hilbert space $\mathcal{H}$ as a product of traces $\tr_i$ over one-spin Hilbert spaces,
\begin{equation}\label{eq:trace1}
\eqalign{
%\Tr\left[\left(\prod_{m=1}^n \prod_{i=1}^N \rme^{-\xi\left(i,\bi{x}^{(m)}\right)/n}\right) \exp\left(-t\sum_{m=1}^N\sigma_m^3\right)\right]\\ \qquad=
\fl\Tr\left[\left(\prod_{m=1}^n \prod_{i=1}^N \rme^{-\xi\left(i,\bi{x}^{(m)}\right)/n}\right) \prod_{m=1}^N\rme^{-t\sigma_m^3}\right]
\sim \Tr\left(\prod_{m=1}^n \prod_{i=1}^N \rme^{-[\xi\left(i,\bi{x}^{(m)}\right)+t\sigma_i^3]/n}\right)\\
= \prod_{i=1}^N \tr_i\left(\prod_{m=1}^n \rme^{-\left[\xi\left(i,\bi{x}^{(m)}\right)+t\sigma_i^3\right]/n}\right)\\
\sim\prod_{i=1}^N \tr_i \left\{\exp\left[\frac{1}{n}\sum_{m=1}^n\left(\sum_{\alpha=1}^3 x^{(m)}_\alpha \sqrt{\lambda_\alpha}\, \sigma_i^\alpha-t\sigma_i^3\right)\right]\right\}.
}
\end{equation}
For the same kind of reasoning as outlined below \eref{eq:replace1}, commutators of order $1/n^2$ have been neglected. In the limit $n\to\infty$ this approximation will become exact and the asymptotic equalities $\sim$ in \eref{eq:trace1} become proper equalities. With the definition
\begin{equation}
c_\alpha=c_\alpha\bigl(\bigl\{x_\alpha^{(m)}\bigr\}\bigr)=\frac{1}{n}\sqrt{\lambda_\alpha}\sum_{m=1}^n x_\alpha^{(m)} - \delta_{\alpha,3}t,
\end{equation}
we can rewrite \eref{eq:trace1} as 
\begin{equation}\label{eq:trace2}
\fl\prod_{i=1}^N \tr\left[\exp\left(\!\!\begin{array}{cc}c_3&c_1-\rmi c_2\\c_1+\rmi c_2&-c_3\end{array}\!\!\right)\right] = \left[\tr\left(\!\!\begin{array}{ll}\rme^{r_t}&0\\0&\rme^{-r_t}\end{array}\!\!\right)\right]^N = \left(2\cosh r_t\right)^N,
\end{equation}
where
\begin{equation}
r_t\equiv r_t\bigl(\bigl\{\bi{x}^{(m)}\bigr\}\bigr) = \sqrt{c_1^2+c_2^2+c_3^2}.
\end{equation}
Replacing the trace in \eref{eq:Omega2} by expression \eref{eq:trace2}, we obtain for the density of states
\begin{equation}\label{eq:Omega3}
\eqalign{
\fl\Omega_N(e,m) = \frac{2^N N^2}{4\pi^2}\int_{a-\rmi\infty}^{a+\rmi\infty}\rmd s \int_{b-\rmi\infty}^{b+\rmi\infty}\rmd t\\
\times\lim_{n\to\infty}\left(\frac{N}{2\pi ns}\right)^{3n/2}\int\!\cdots\!\int\rmd^3 x^{(1)}\cdots\rmd^3 x^{(n)} \rme^{N\mathcal{F}(s,t,\{\bi{x}^{(m)}\})},
}
\end{equation}
where
\begin{equation}\label{eq:F}
\fl\mathcal{F}(s,t,\{\bi{x}^{(m)}\})=es+mt-\frac{1}{2ns}\sum_{m=1}^n \bi{x}^{(m)}\cdot\bi{x}^{(m)}+\ln\cosh \bigl[r_t\bigl(\bigl\{\bi{x}^{(m)}\bigr\}\bigr)\bigr].
\end{equation}

\subsubsection{Asymptotic evaluation of the $(3n+2)$-dimensional integral.} We have to deal with a $(3n+2)$-dimensional integral, to be evaluated in the limit $n\to\infty$ and, afterwards, the thermodynamic limit $N\to\infty$. This integral \eref{eq:Omega3} is of Laplace-type with respect to the large parameter $N$, and an asymptotic evaluation can be performed by a multidimensional version of the method of steepest descent (see \cite{Miller} for a textbook presentation).

To apply this method, we need to find a stationary point of the function $\mathcal{F}$ for which it is possible to smoothly deform the contours of the $s$- and $t$-integrations such that the paths of integration correspond to constant (zero) imaginary part of $\mathcal{F}$. Stationary points of $\mathcal{F}$ need to satisfy the conditions
\numparts
\begin{eqnarray}
0&=\frac{\partial\mathcal{F}}{\partial s} = e+\frac{1}{2ns^2}\sum_{m=1}^n\bi{x}^{(m)}\cdot\bi{x}^{(m)},\\
0&=\frac{\partial\mathcal{F}}{\partial t} = m+\frac{\tanh r_t}{r_t}\left(t-\frac{\sqrt{\lambda_3}}{n}\sum_{m=1}^n x_3^{(m)}\right),\label{eq:saddle1b}\\
0&=\frac{\partial\mathcal{F}}{\partial x_\alpha^{(u)}} = -\frac{x_\alpha^{(u)}}{ns}-\frac{\tanh r_t}{r_t} \frac{\sqrt{\lambda_\alpha}}{n}\left(\delta_{\alpha,3}t-\frac{\sqrt{\lambda_\alpha}}{n}\sum_{m=1}^n x_\alpha^{(m)}\right),\label{eq:saddle1c}
\end{eqnarray}
\endnumparts
where $\alpha\in\{1,2,3\}$ and $u\in\{1,\dots,n\}$. Inserting \eref{eq:saddle1b} in \eref{eq:saddle1c} and some straight\-for\-ward algebra allows us to rewrite this set of equations as
\numparts
\begin{eqnarray}
0&= 2nes^2+\sum_{m=1}^n\bi{x}^{(m)}\cdot\bi{x}^{(m)},\\
0&= mr_t+\left(t-\frac{\sqrt{\lambda_3}}{n}\sum_{m=1}^n x_3^{(m)}\right)\tanh r_t,\label{eq:saddle2b}\\
0&= x_3^{(u)}-ms\sqrt{\lambda_3},\label{eq:saddle2c}\\
0&= x_\alpha^{(u)}\left(t-\frac{\sqrt{\lambda_3}}{n}\sum_{m=1}^n x_3^{(m)}\right)+\frac{ms\lambda_\alpha}{n}\sum_{m=1}^n x_\alpha^{(m)},
\end{eqnarray}
\endnumparts
where $\alpha\in\{1,2\}$ and $u\in\{1,\dots,n\}$. There is a class of particularly simple solutions to this set of equations where all the $\bi{x}^{(m)}$ are identical, i.e.
\begin{equation}\label{eq:xequal}
\bi{x}^{(m)} = \bi{x} = (x_1,x_2,x_3)\qquad\forall m\in\{1,\dots,n\}.
\end{equation}
Similar to the canonical calculation reported in \cite{TinCa74}, it should be possible to prove that one of the stationary points of $\mathcal{F}$ which is subject to condition \eref{eq:xequal} corresponds indeed to the maximum of the exponent along the properly deformed integration path and therefore yields the correct result for the integral \eref{eq:Omega3} in the thermodynamic limit. We have not gone through this calculation explicitely, but the final result, especially when compared to the results of \sref{sec:special}, seems to confirm this assumption beyond any reasonable doubt.  Under the assumption \eref{eq:xequal}, the set of equations simplifies to
\numparts
\begin{eqnarray}
0&= 2es^2+\bi{x}^2,\label{eq:saddle3a}\\
0&= mR_t(\bi{x})+\left(t-\sqrt{\lambda_3}x_3\right)\tanh R_t(\bi{x}),\label{eq:saddle3b}\\
0&= x_3-ms\sqrt{\lambda_3},\label{eq:saddle3c}\\
0&= x_\alpha\left[ms\left(\lambda_3-\lambda_\alpha\right)-t\right],\qquad\alpha\in\{1,2\},\label{eq:saddle3d}
\end{eqnarray}
\endnumparts
where
\begin{equation}\label{eq:Rt1}
R_t(\bi{x}) = \sqrt{\lambda_1 x_1^2+\lambda_2 x_2^2+\left(t-\sqrt{\lambda_3} x_3\right)^2}.% = \sqrt{\lambda_1 x_1^2+\lambda_2 x_2^2+\left(t-ms\lambda_3\right)^2}.
\end{equation}
%where \eref{eq:saddle3c} has been used.
For an asymptotic evaluation of the integrals in \eref{eq:Omega3} by means of the method of steepest descent, we have to evaluate $\mathcal{F}(s,t,\{\bi{x}^{(m)}\})$ as defined in \eref{eq:F} at the values of $s$, $t$, and $\{\bi{x}^{(m)}\}$ specified by \eref{eq:saddle3a}--\eref{eq:saddle3d}. For $\mathcal{F}$ we obtain under condition \eref{eq:xequal} the expression
\begin{equation}\label{eq:F2}
\mathcal{F}(s,t,\bi{x})=es+mt-\frac{\bi{x}^2}{2s}-\frac{1}{2}\ln\left[1-\tanh^2 R_t(\bi{x})\right],
\end{equation}
where the identity $2\ln\cosh x=-\ln(1-\tanh^2x)$ has been used. Making use of the identities \eref{eq:saddle3a}--\eref{eq:saddle3d}, it is a matter of straightforward algebra to evaluate $\mathcal{F}$ at the values $s_0$, $t_0$, and $\bi{x}_0$ which are solutions of this set of equations %. Depending on whether $\lambda_1$ is larger than $\lambda_2$ or not, the relevant stationary point of $\mathcal{F}$ satisfies either $x_1=0$ and $t=ms(\lambda_3-\lambda_2)$, or $x_2=0$ and $t=ms(\lambda_3-\lambda_1)$ 
(see \ref{sec:evalF} for the details of the calculation). Evaluating $\mathcal{F}$ at these points and taking into account \eref{eq:saddle3a}--\eref{eq:saddle3c}, we obtain
\begin{equation}\label{eq:Ffinal}
\fl -2\mathcal{F}(s_0,t_0,\bi{x}_0)=[1-f(e,m)]\ln[1-f(e,m)]+[1+f(e,m)]\ln[1+f(e,m)]
\end{equation}
with
\begin{equation}\label{eq:fem}
f(e,m)=\sqrt{m^2\left(1-\frac{\lambda_3}{\lambda_\perp}\right)-\frac{2e}{\lambda_\perp}}
\end{equation}
as defined previously in \eref{eq:femspecial}. The constant $\lambda_\perp$ can be either $\lambda_1$ or $\lambda_2$, unless one of them vanishes. The solution relevant for the asymptotic evaluation of the integral \eref{eq:Omega3} is the one which maximizes $\mathcal{F}$. It is straightforward to verify that this solution corresponds to\footnote{The case of $\lambda_1=0=\lambda_2$ has to be treated separately, but this is just the well-known case of the Curie-Weiss Ising model.}
\begin{equation}\label{eq:lambdaperp}
\lambda_\perp=\max\{\lambda_1,\lambda_2\}.
\end{equation}
Solutions of the type \eref{eq:Ffinal} exist for all values of $(e,m)$ which satisfy the inequalities
\begin{equation}
2e+m^2\lambda_3<0\qquad\mbox{and}\qquad \lambda_\perp > m^2\left(\lambda_\perp-\lambda_3\right)-2e
\end{equation}
(see again \ref{sec:evalF} for a derivation).

According to the method of steepest descent, the asymptotic behaviour of $\Omega_N$ in \eref{eq:Omega3} is now given as $\exp[N(\ln2+\mathcal{F})]$ times some prefactor (see for example section 3.7 of Miller's textbook \cite{Miller} for the prefactor of multidimensional Laplace integrals, which can be adapted to the method of steepest descent of multidimensional integrals). The prefactor, however, is subexponential in $N$. Since we are interested in the microcanonical entropy in the thermodynamic limit,
\begin{equation}
s(e,m)=\lim_{N\to\infty}\frac{1}{N}\ln\Omega_N(e,m),
\end{equation}
subexponential terms do not contribute, and we obtain the following final result for the microcanonical entropy of the anisotropic quantum Heisenberg model in the thermodynamic limit:
\begin{equation}\label{eq:sfinal}
\fl s(e,m)=\ln2-\frac{1}{2}[1-f(e,m)]\ln[1-f(e,m)]-\frac{1}{2}[1+f(e,m)]\ln[1+f(e,m)],
\end{equation}
defined on the domain
\begin{equation}\label{eq:domain}
\fl\mathcal{D}=\left\{(e,m)\in\RR^2\,\big|\,2e+m^2\lambda_3<0\;\mbox{and}\;\lambda_\perp > m^2\left(\lambda_\perp-\lambda_3\right)-2e\right\},
\end{equation}
where $f(e,m)$ and $\lambda_\perp$ are specified in \eref{eq:fem} and \eref{eq:lambdaperp}.

\begin{figure}\center
%\vspace{1mm}
\parbox{3cm}{\includegraphics[width=26mm]{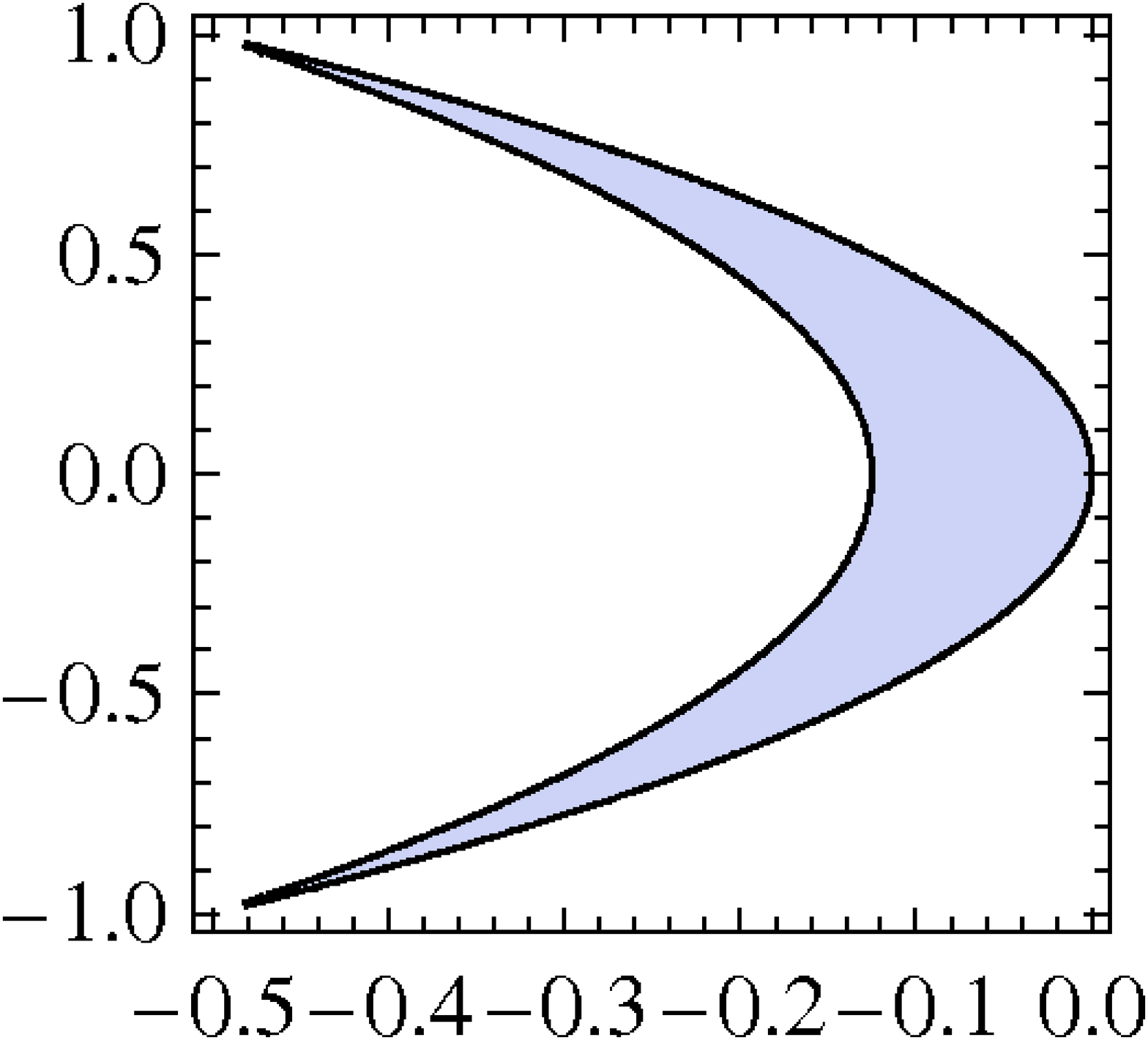}}
\hspace{12mm}
\parbox{4cm}{\includegraphics[width=33mm]{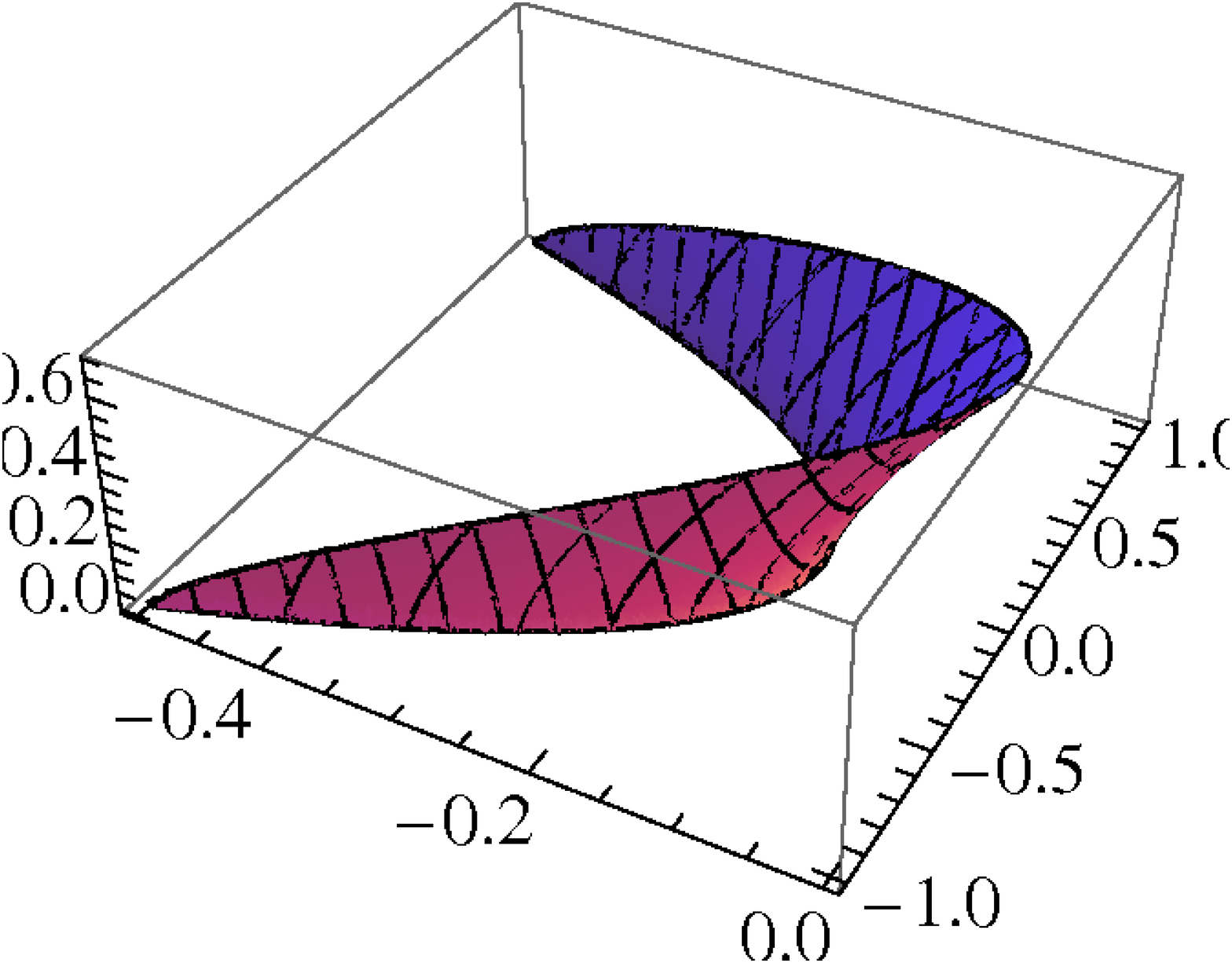}}
\newline
\parbox{3cm}{\includegraphics[width=26mm]{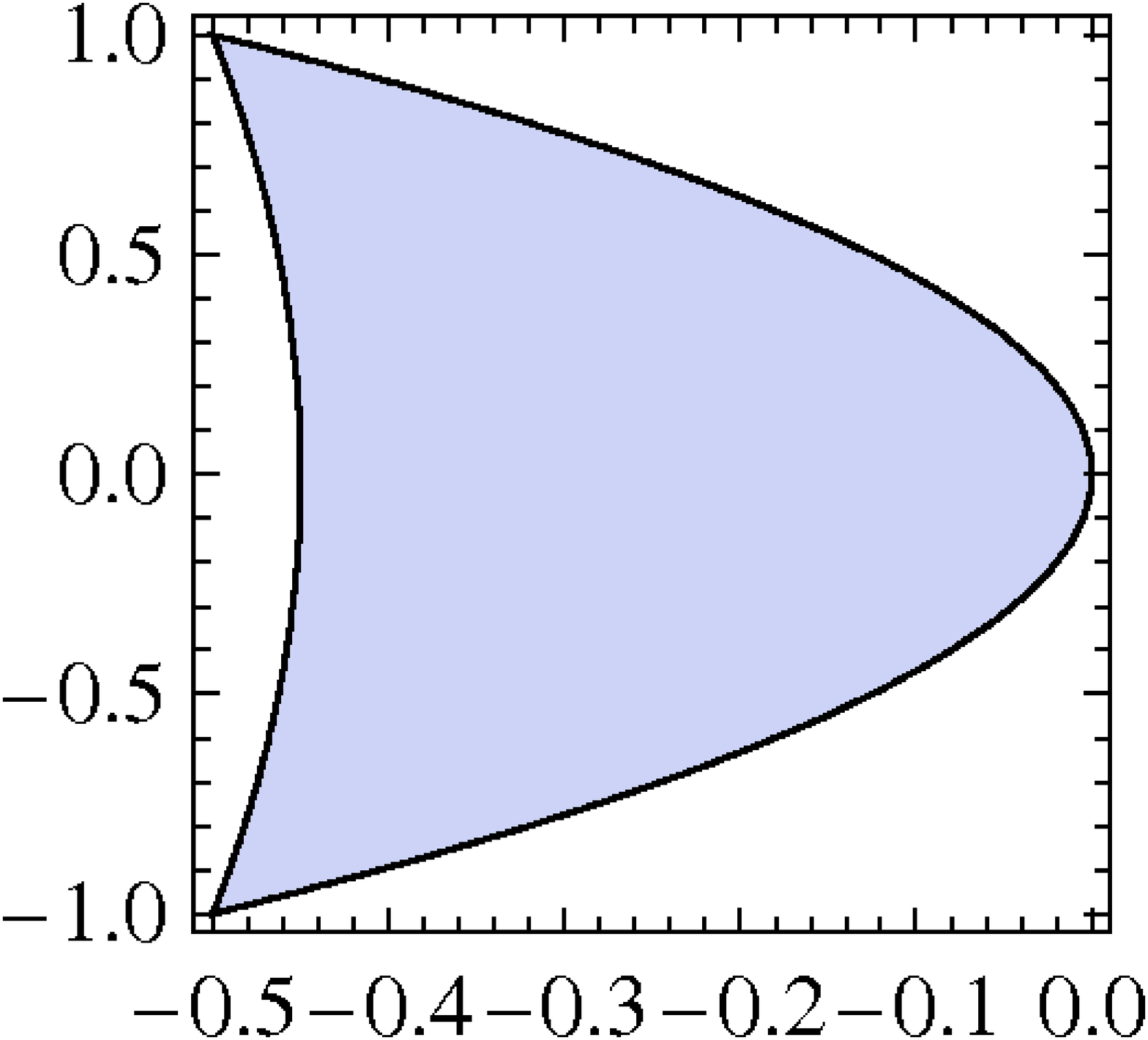}}
\hspace{12mm}
\parbox{4cm}{\includegraphics[width=33mm]{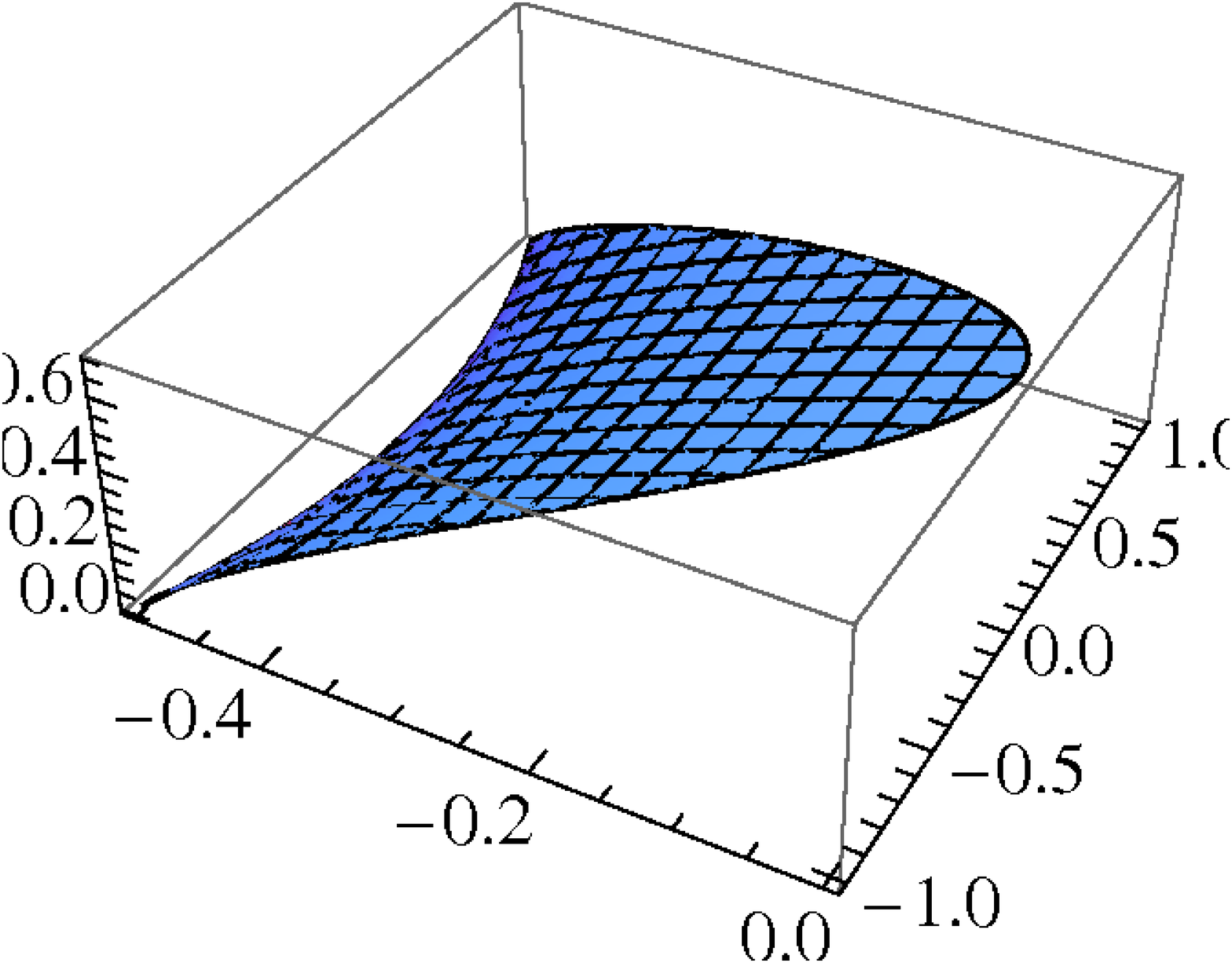}}
\newline
\parbox{3cm}{\includegraphics[width=26mm]{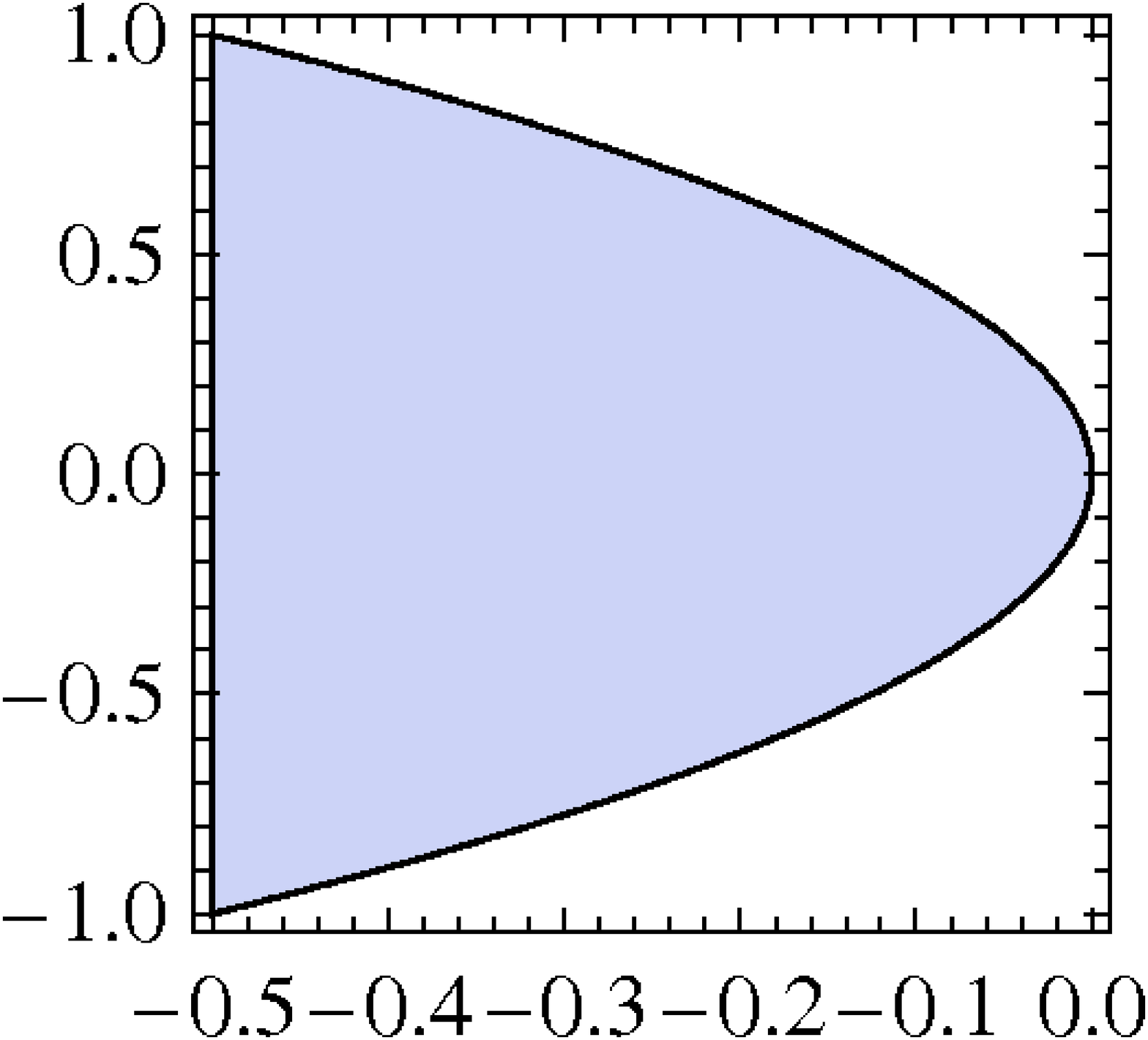}}
\hspace{12mm}
\parbox{4cm}{\includegraphics[width=33mm]{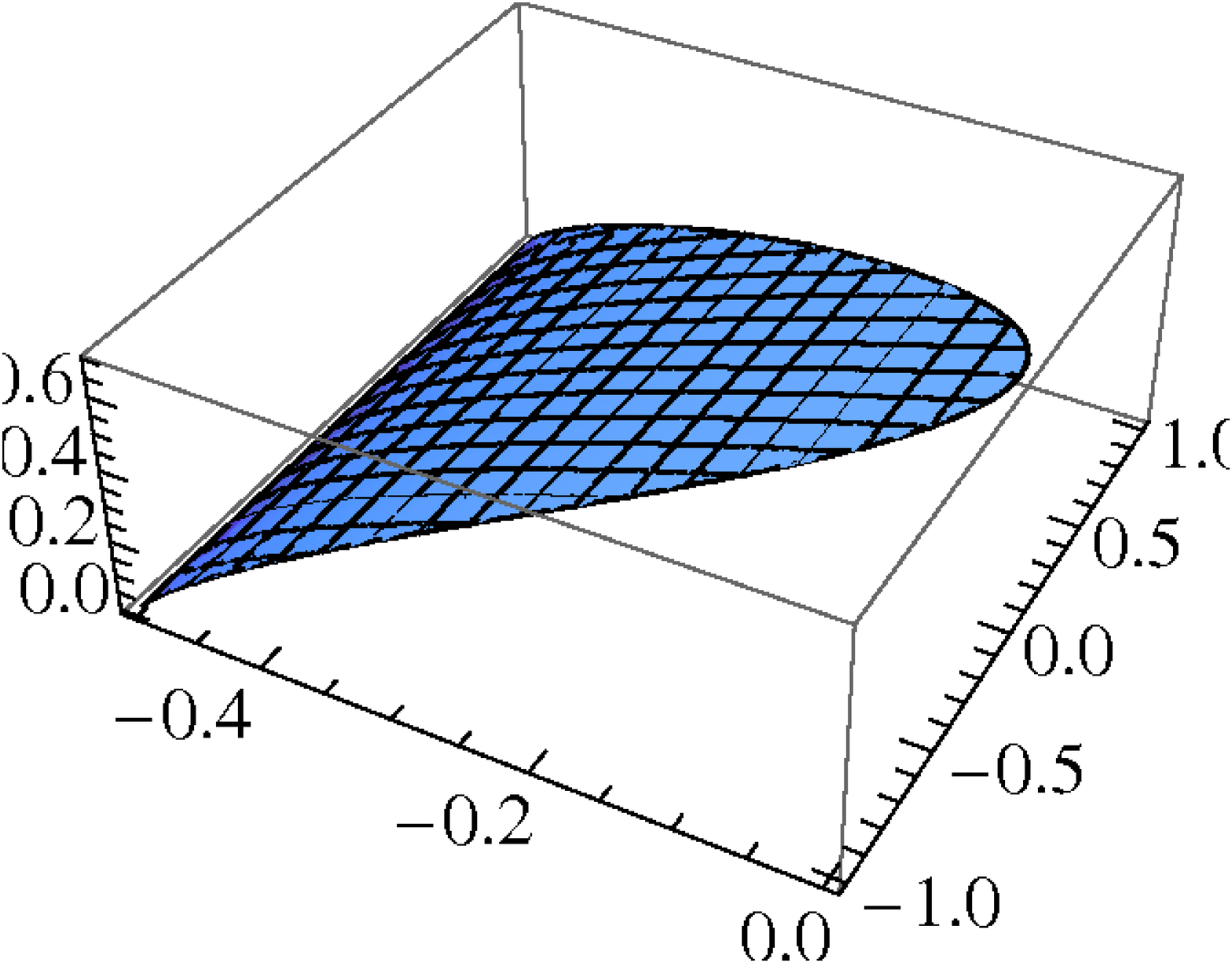}}
\newline
\parbox{3cm}{\includegraphics[width=26mm]{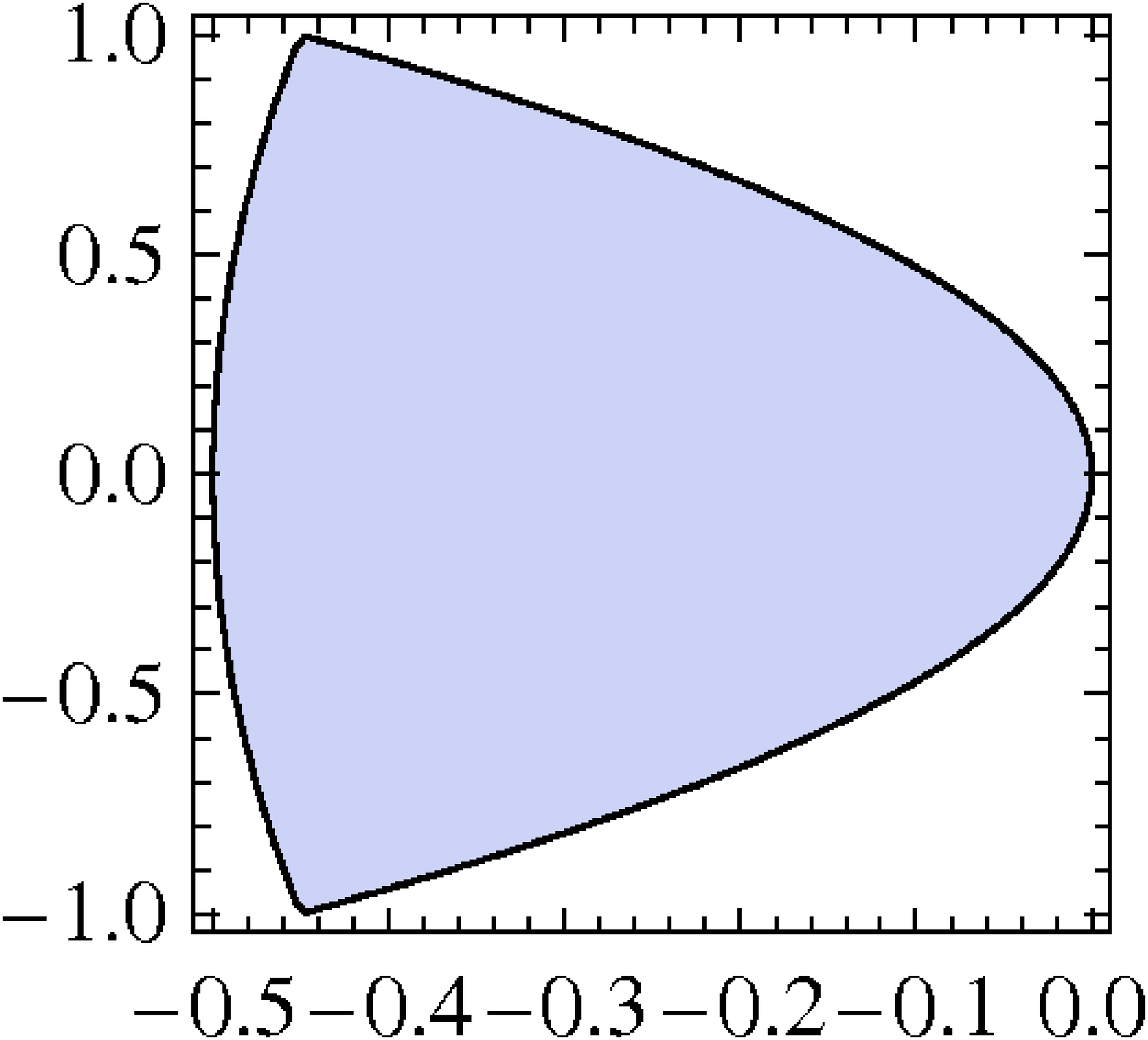}}
\hspace{12mm}
\parbox{4cm}{\includegraphics[width=33mm]{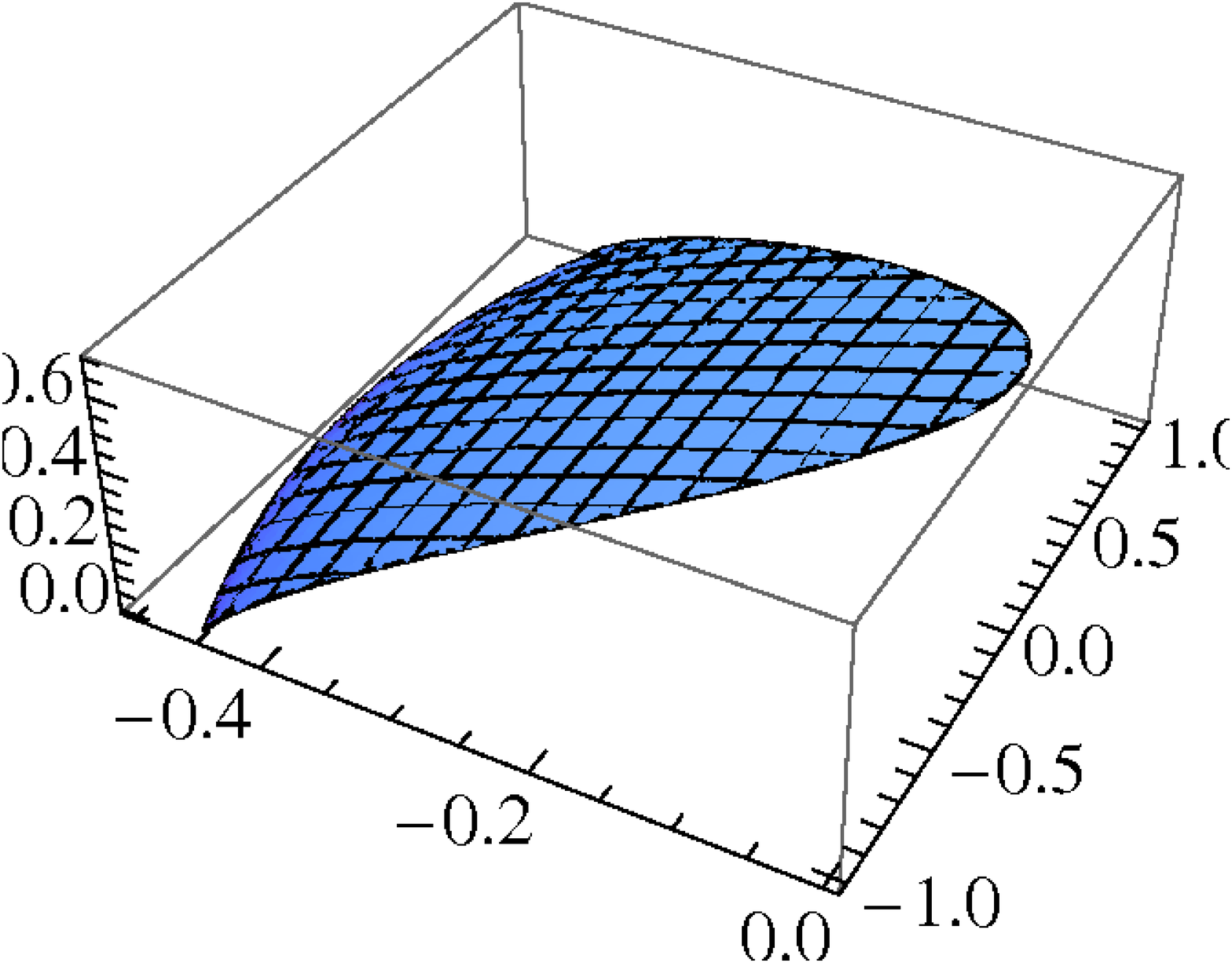}}
\newline
\parbox{3cm}{\includegraphics[width=26mm]{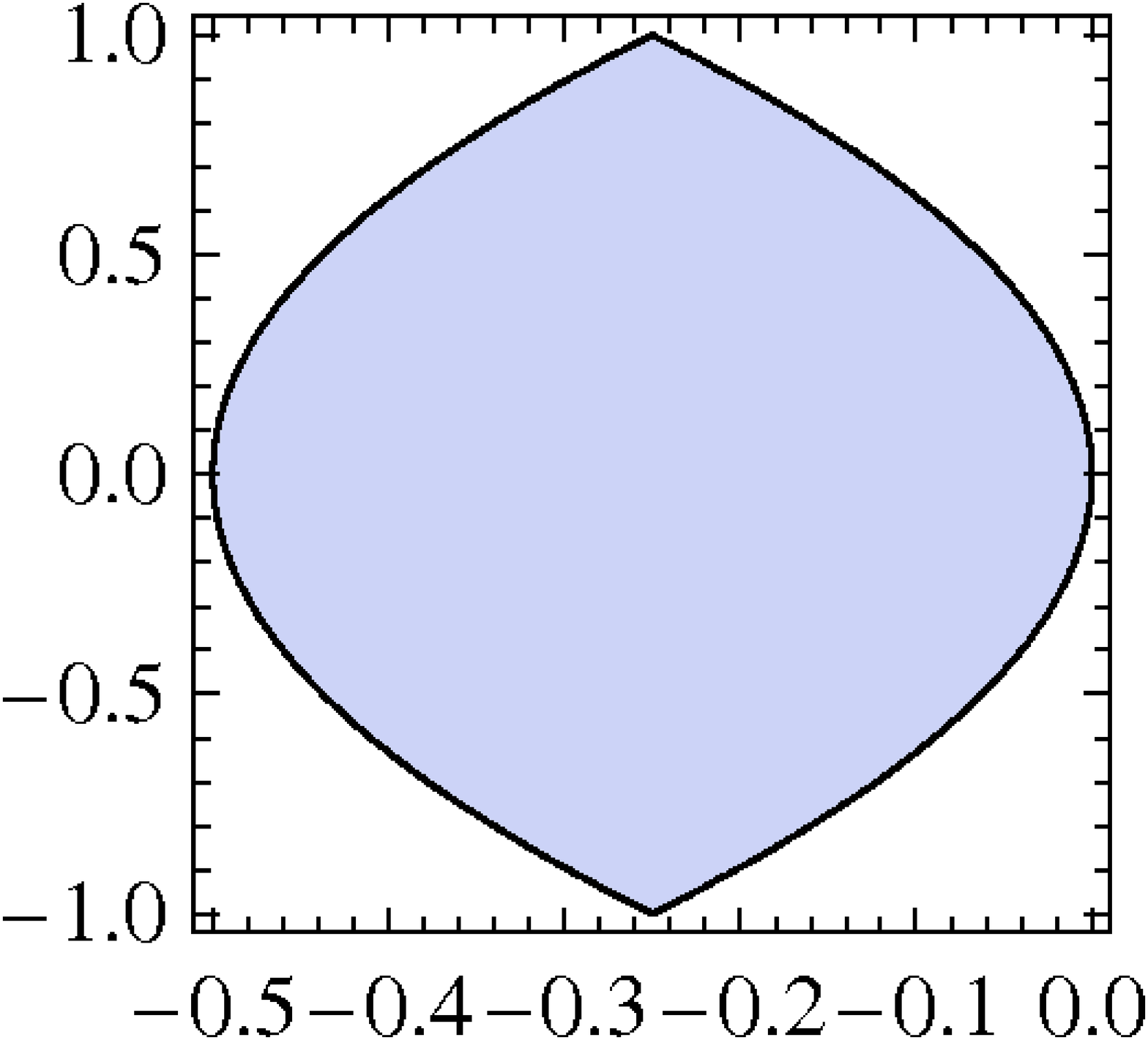}}
\hspace{12mm}
\parbox{4cm}{\includegraphics[width=33mm]{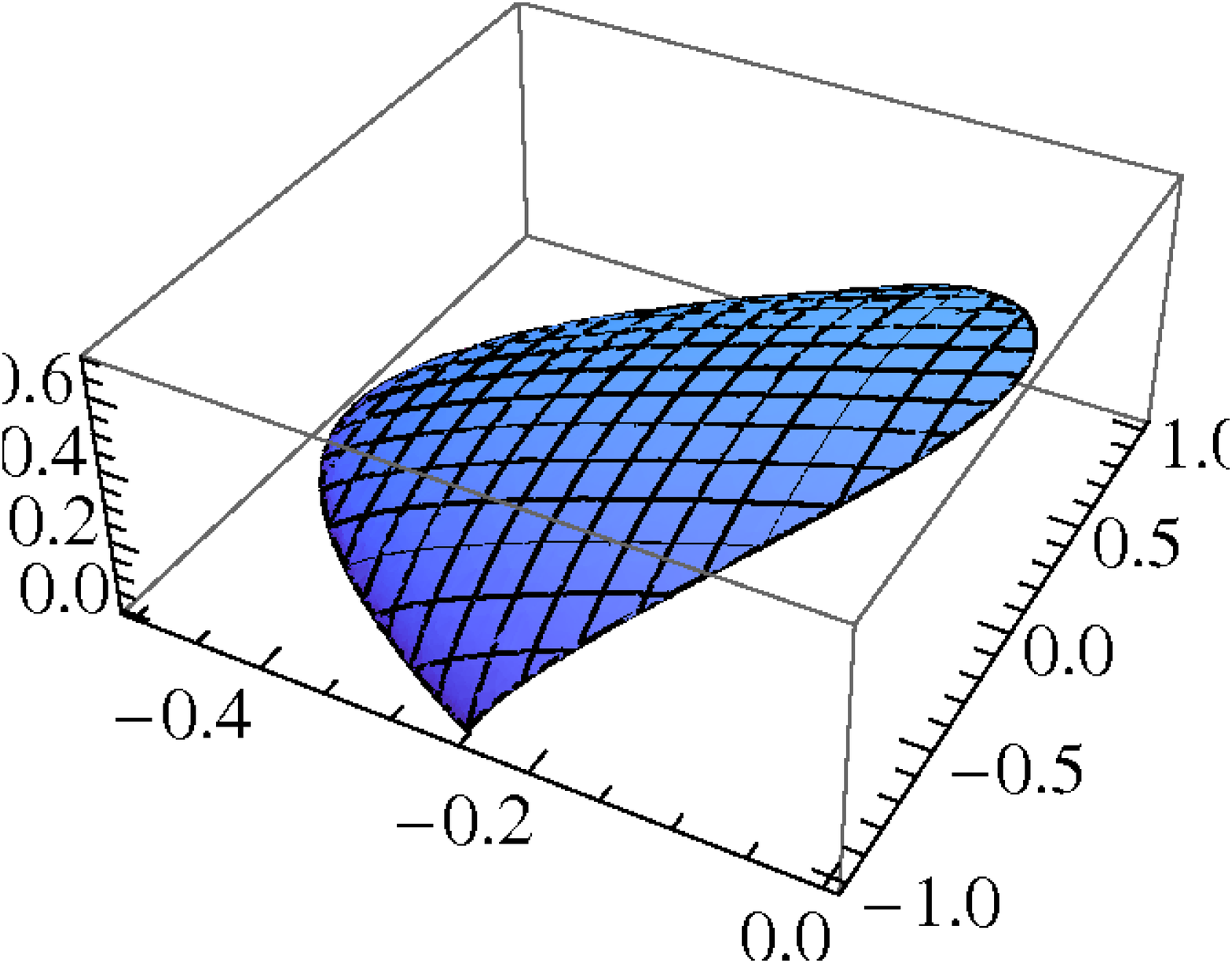}}
\newline
\parbox{3cm}{\includegraphics[width=26mm]{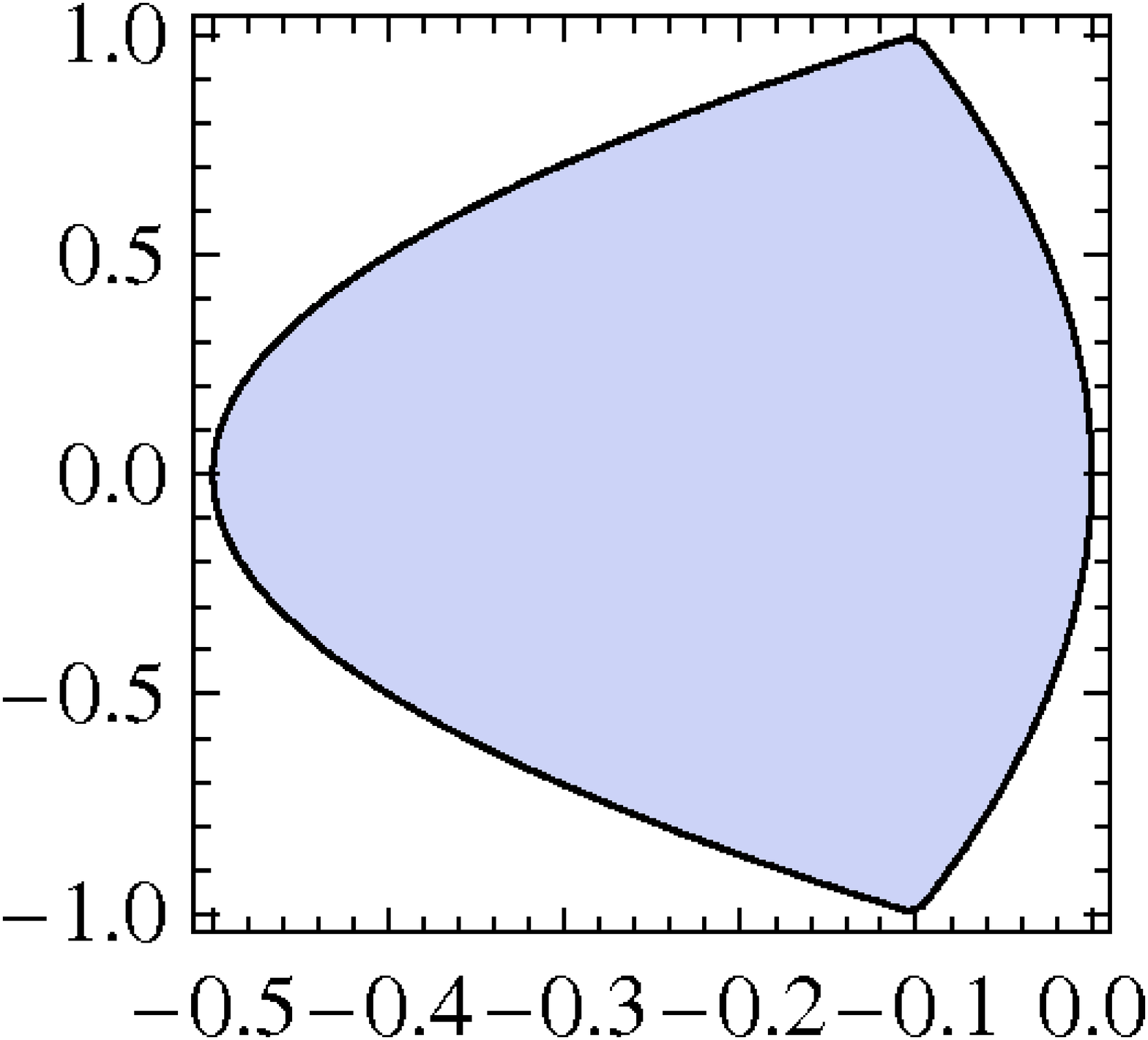}}
\hspace{12mm}
\parbox{4cm}{\includegraphics[width=33mm]{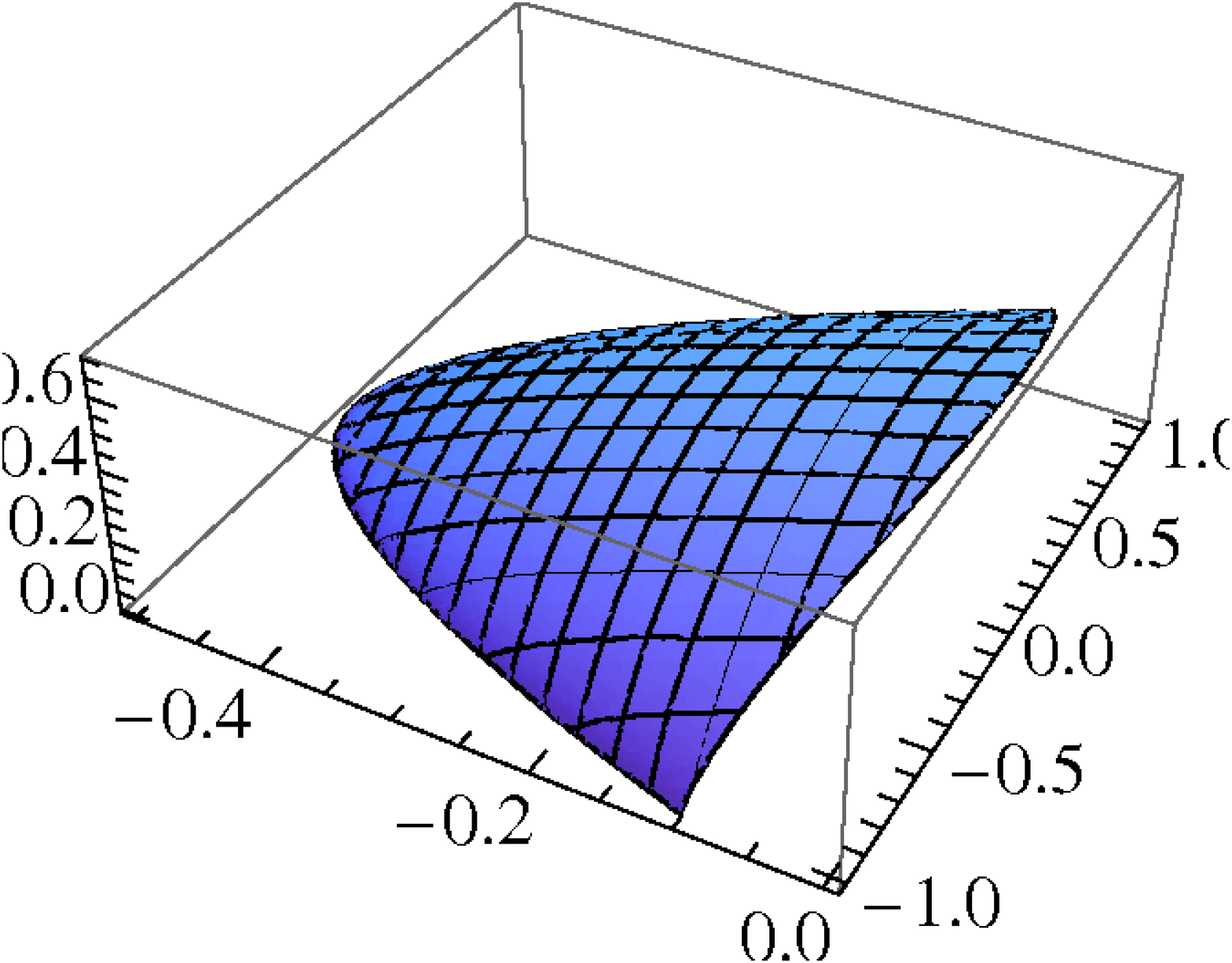}}
\newline
\parbox{3cm}{\includegraphics[width=26mm]{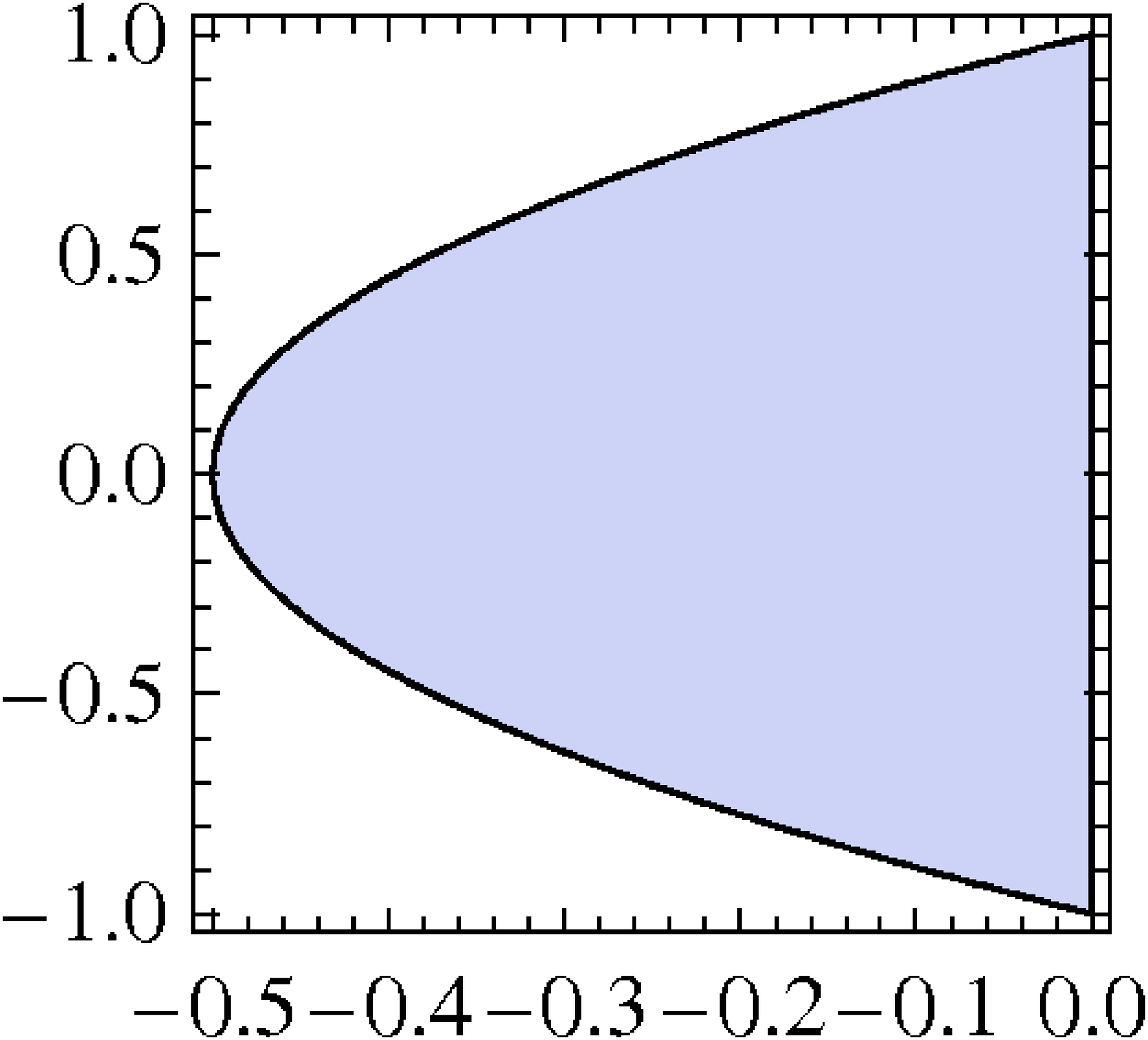}}
\hspace{12mm}
\parbox{4cm}{\includegraphics[width=33mm]{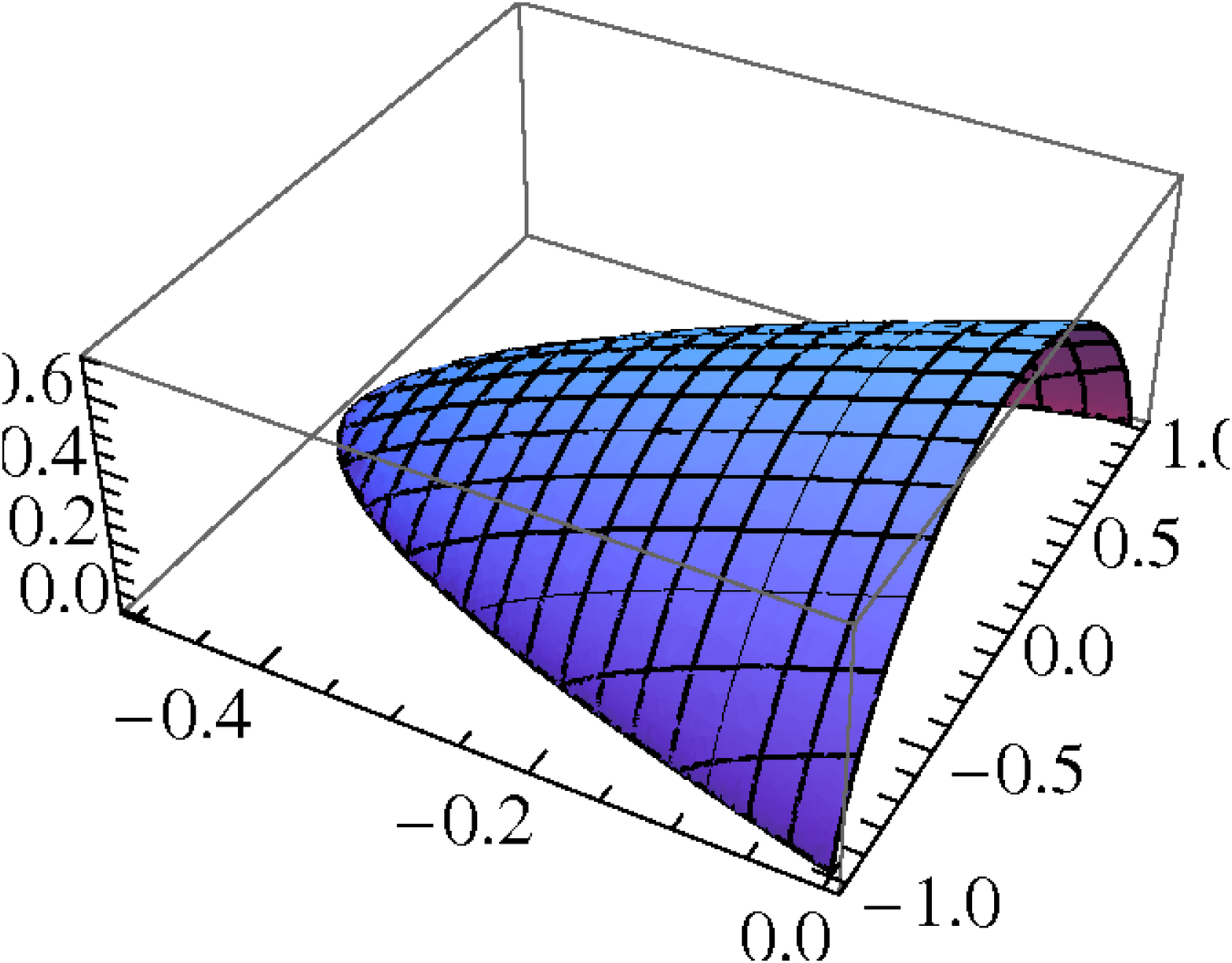}}
\newline
\caption{\label{fig:entropy}
Domains $\mathcal{D}$ (left) and graphs (right) of the microcanonical entropy $s(e,m)$ of the quantum Heisenberg model for some combinations of the coupling $\lambda_\perp,\lambda_3$. From top to bottom: $(\lambda_\perp,\lambda_3)=(1/4,1)$, $(9/10,1)$, $(1,1)$, $(1,9/10)$, $(1,1/2)$, $(1,1/5)$, $(1,0)$. For the domains, the abscissa is the energy $e$ and the ordinate is the magnetization $m$, and the entropy is defined on the shaded area.
}
\end{figure}

This result is remarkably simple, in the sense that an explicit expression for $s(e,m)$ can be given. This is in contrast to the canonical ensemble, where the canonical free energy $g(\beta,h)$ is given implicitly as the solution of a maximization (see \sref{sec:Legendre} for more details on the canonical solution). It may seem a bit disappointing that the solution \eref{eq:sfinal} is identical to the one we obtained already from the much simpler calculation for the special case $\lambda_1=\lambda_2$ in \sref{sec:special}, with the only difference that now $\lambda_\perp$ is defined according to equation \eref{eq:lambdaperp}.
Plots of the domains and graphs of $s(e,m)$ are shown in \fref{fig:entropy} for a number of coupling strengths $\lambda_\perp$, $\lambda_3$.
%------------------------------------------------
%------------------------------------------------

%----------------------------------------------------
\section{Recovering the canonical Gibbs free energy}
\label{sec:Legendre}
%----------------------------------------------------

The microcanonical entropy we have computed in \sref{sec:micent} forms the starting point of an analysis of the anisotropic quantum Heisenberg model in the microcanonical ensemble. An analogous role is played by the canonical Gibbs free energy 
\begin{equation}
g_N(\beta,h)=-\frac{1}{N\beta}\ln\Tr\rme^{-\beta H_h}
\end{equation}
for calculations in the canonical ensemble. In the thermodynamic limit, the corresponding infinite-system quantity
\begin{equation}
g(\beta,h)=\lim_{N\to\infty}g_N(\beta,h)
\end{equation}
is related to the microcanonical entropy $s(\varepsilon,m)$ by means of a Legendre-Fenchel transform,
\begin{equation}\label{eq:LegendreFenchelDef}
-\beta g(\beta,h)=\sup_{e,m}\left[s(e,m)-\beta e+\beta hm\right].
\end{equation}
Note that while, in the thermodynamic limit, the canonical free energy is always given as the Legendre transform of the microcanonical entropy, the inverse is in general not true. In particular, it would have been impossible to derive the entropy \eref{eq:sfinal} from the canonical Gibbs free energy for anisotropy parameters $\lambda_\perp<\lambda_3$ for which the entropy is nonconcave. This will be discussed in more detail in \sref{sec:nonequivalence}.

In order to determine $g$, it is helpful to consider the following operational interpretation of the supremum in \eref{eq:LegendreFenchelDef}: Take a plane described by the equation
\begin{equation}\label{eq:plane}
\bar{s}_{\beta,h,c}(e,m)=c+\beta e-\beta hm,
\end{equation}
i.e.\ with slopes $\beta$ in $e$-direction and $-\beta h$ in $m$-direction. Start with a very large value of the parameter $c$ and lower this value, and therefore the plane, until, at some value of $c(\beta,h)$, the plane touches the graph of $s(e,m)$ for the first time. Then the value of $g(\beta,h)$ is related to $\bar{s}_{\beta,h,c(\beta,h)}$ evaluated at the origin,
\begin{equation}\label{eq:gfromsbar}
-\beta g(\beta,h)=\bar{s}_{\beta,h,c(\beta,h)}(0,0).
\end{equation}
More details on this graphic-geometric interpretation, although for the special case of a Legendre (not Legendre-Fenchel) transform, can be found in section III.A and figure 3 of \cite{Zia_etal09}. We will now use this interpretation to calculate the Legendre-Fenchel transform of $s$, separately for the cases $\lambda_\perp<\lambda_3$ and $\lambda_\perp>\lambda_3$.

\subsection{$\lambda_\perp<\lambda_3$}
\label{sec:nonconcave}

\begin{figure}\center
\includegraphics[width=55mm]{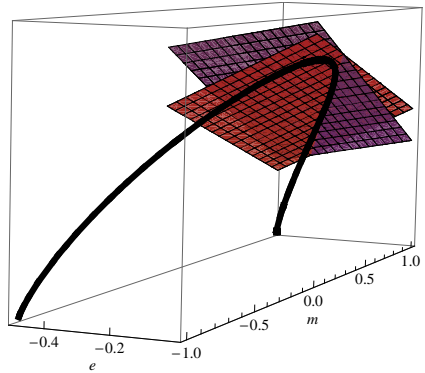}
\caption{\label{fig:planes}
Plot of the boundary curve described by \eref{eq:eofm} and \eref{eq:sofm} together with two examples of tangent planes touching the curve at the point $(e(m_0),m_0,s(m_0))$ with $m_0=1/2$. The planes have slopes $\beta=1/5$, respectively $\beta=1/2$, in the $e$-direction, while their slopes in $m$-direction are determined by \eref{eq:constraint}.
}
\end{figure}

\begin{figure}\center
\psfrag{h}{\footnotesize $h$}
\psfrag{bl3}{\footnotesize $\beta\lambda_3$}
\psfrag{hl1l3l1l3}{\footnotesize $h/(\lambda_\perp-\lambda_3)$}
\psfrag{bl1}{\footnotesize $\beta\lambda_\perp$}
\parbox{5cm}{\includegraphics[width=50mm]{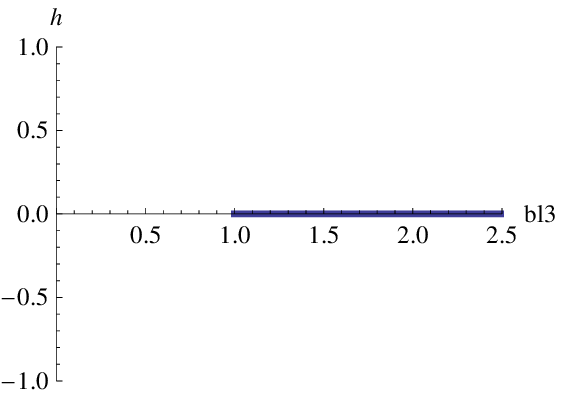}}
\hspace{12mm}
\parbox{5cm}{\includegraphics[width=50mm]{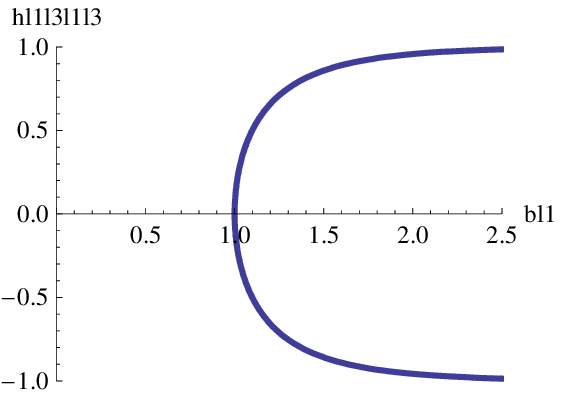}}
\caption{\label{fig:canPD}
Canonical phase diagram of the Curie-Weiss anisotropic quantum Heisenberg model. The lines in the (rescaled) $(\beta,h)$-planes indicate the values of inverse temperature and magnetic field at which the canonical Gibbs free energy $g(\beta,h)$ is nonanalytic. Left: Phase diagram for anisotropy parameters $\lambda_\perp<\lambda_3$. For small values of $\beta\lambda_3$, the system is in a paramagnetic phase, whereas it becomes ferromagnetic for larger values. Right: Phase diagram for anisotropy parameters $\lambda_\perp>\lambda_3$. Inside the $\subset$-shaped curve, a non-zero magnetization in the $1$-$2$-plane occurs, whereas outside this area the magnetization vector points into the $3$-direction.
}
\end{figure}

By inspection of rows one and two of \fref{fig:entropy} [or by analysis of the results in \eref{eq:sfinal} and \eref{eq:domain}], one can convince oneself that, for $\lambda_\perp<\lambda_3$, a plane of the type \eref{eq:plane} when lowered onto the graph of $s$ will always touch the graph at a boundary point of the domain $\mathcal{D}$ of $s$, given by
\begin{equation}\label{eq:eofm}
e(m)=-\textstyle{\frac{1}{2}}m^2\lambda_3,
\end{equation}
where
\begin{equation}\label{eq:sofm}
\fl s(m)\equiv s\left(-\case{1}{2}m^2\lambda_3,m\right)=\ln2-\case{1}{2}(1-m)\ln(1-m)-\case{1}{2}(1+m)\ln(1+m).
\end{equation}
Next we need to construct the family of all planes tangent to this boundary curve at a given point $m=m_0$; see \fref{fig:planes} for an illustration.
%------------------------------------------------
%------------------------------------------------
A tangent vector to the boundary curve is
\begin{equation}
\fl \bi{t}(m_0)=\hat{\bi{e}}\partial_{m_0}e(m_0)+\hat{\bi{m}}\partial_{m_0}m_0+\hat{\bi{s}}\partial_{m_0}s(m_0)=-\hat{\bi{e}}m_0\lambda_3+\hat{\bi{m}}-\hat{\bi{s}}\,\arctanh(m_0),
\end{equation}
where $\hat{\bi{e}}$, $\hat{\bi{m}}$, and $\hat{\bi{s}}$ are unit vectors in $e$, $m$, and $s$ direction. A family of vectors $\bi{n}(m_0)$ which are perpendicular to $\bi{t}(m_0)$ can be constructed by defining
\begin{equation}\label{eq:normal}
\bi{n}(m_0)=-\beta\hat{\bi{e}}+\beta h\hat{\bi{m}}+\hat{\bi{s}}
\end{equation}
and requiring that
\begin{equation}\label{eq:constraint}
\bi{n}(m_0)\cdot\bi{t}(m_0) = \beta m_0\lambda_3+\beta h-\arctanh(m_0)=0.
\end{equation}
The constraint \eref{eq:constraint} fixes one of the parameters $\beta$ and $h$ in \eref{eq:normal}, while the other parameter labels a family of (non-normalized) vectors. To obtain the family of tangent planes to the boundary curve, we consider each of the vectors $\bi{n}(m_0)$ as a normal vector of such a plane. We further demand that each plane touches the curve in the point $\bi{p}(m_0)=\left(e(m_0),m_0,s(m_0)\right)$. All points $\bi{r}=(e,m,s)$ lying in this plane have to fulfill the equation
\begin{equation}
\fl 0=\bi{n}(m_0)\cdot\left(\bi{r}-\bi{p}(m_0)\right)=-\beta\left(e+\case{1}{2}m_0^2\lambda_3\right)+\beta h(m-m_0)+s-s(m_0).
\end{equation}
Therefore, the planes $\bar{s}$ touching the graph of $s$ as explained above are given by
\begin{equation}
\bar{s}_{\beta,h}=s(m_0)+\beta\left(e+\case{1}{2}m_0^2\lambda_3\right)-\beta h(m-m_0),
\end{equation}
where $\beta$, $h$, and $m_0$ are subject to \eref{eq:constraint}. For $\beta\lambda_3<1$ or for sufficiently large magnetic fields $h$, equation \eref{eq:constraint} has just a single solution, whereas for $\beta\lambda_3>1$ and small enough $h$ three solutions exist. In the latter case, the relevant plane $\bar{s}$ corresponding to the supremum in \eref{eq:LegendreFenchelDef} is given by the $m_0$ with the largest absolute value. According to \eref{eq:gfromsbar} we have
\begin{equation}
-\beta g(\beta,h)=s(m_0)+\case{1}{2}\beta m_0^2\lambda_3+\beta h m_0
\end{equation}
and, making use of \eref{eq:constraint}, this result can be written as
\begin{equation}\label{eq:gfinal}
-\beta g(\beta,h)=\ln2-\case{1}{2}\ln\left(1-m_0^2\right)-\case{1}{2}\beta m_0^2\lambda_3,
\end{equation}
where $m_0\equiv m_0(\beta,h)$ is determined implicitly as the one solution of \eref{eq:constraint} having the largest absolute value. The swapping from one solution branch to another at $h=0$ for inverse temperatures $\beta\lambda_3>1$ gives rise to nonanalytic behaviour of the canonical Gibbs free energy density, resulting in the phase diagram as plotted in \fref{fig:canPD} (left).
%------------------------------------------------
%------------------------------------------------

Note that, in contrast to the explicit expression for the microcanonical entropy \eref{eq:sfinal}, the transcendental equation \eref{eq:constraint} cannot be solved explicitely for $m_0$, and therefore only an implicit expression for $g(\beta,h)$ can be given.

\subsection{$\lambda_\perp>\lambda_3$}

By inspection of rows four to seven in \fref{fig:entropy}, one can infer that, for anisotropy parameters $\lambda_\perp>\lambda_3$, there is a certain range of slopes $-\beta$ and $\beta h$ for which a plane of type \eref{eq:plane} touches the graph of $s(e,m)$ not at a boundary point of the domain $\mathcal{D}$ of $s$ as in \sref{sec:nonconcave}, but in its interior. This is true for precisely those values of $\beta$ and $h$ for which the equations
\numparts
\begin{eqnarray}
\beta = \frac{\partial s(e,m)}{\partial e} = \frac{\arctanh[f(e,m)]}{\lambda_\perp f(e,m)},\\
-\beta h = \frac{\partial s(e,m)}{\partial m} = m\left(\frac{\lambda_3}{\lambda_\perp}-1\right)\frac{\arctanh[f(e,m)]}{f(e,m)},
\end{eqnarray}
\endnumparts
have solutions in $\mathcal{D}$. This set of equations can be rewritten as
\numparts
\begin{eqnarray}
h=(\lambda_\perp-\lambda_3)m,\label{eq:emvonhb1}\\
\beta\lambda_\perp\sqrt{\frac{1}{\lambda_\perp}\left(\frac{h^2}{\lambda_\perp-\lambda_3}-2e\right)}=\arctanh\sqrt{\frac{1}{\lambda_\perp}\left(\frac{h^2}{\lambda_\perp-\lambda_3}-2e\right)},\label{eq:emvonhb2}
\end{eqnarray}
\endnumparts
and one can verify that, for $\lambda_\perp>\lambda_3$, the equations have solutions with $(e,m)\in\mathcal{D}$ for all $\beta>0$ and $h$ satisfying
\begin{equation}\label{eq:bhinequality}
\beta \geqslant \frac{\lambda_\perp-\lambda_3}{h\lambda_\perp}\,\arctanh\left(\frac{h}{\lambda_\perp-\lambda_3}\right).
\end{equation}

For these values of $\beta$ and $h$, the Legendre-Fenchel transform \eref{eq:LegendreFenchelDef} reduces to the more familiar Legendre transform
\begin{equation}
-\beta g(\beta,h)=s(e(\beta,h),m(\beta,h))-\beta e(\beta,h)+\beta hm(\beta,h),
\end{equation}
where $e(\beta,h)$ and $m(\beta,h)$ are solutions of equations \eref{eq:emvonhb1} and \eref{eq:emvonhb2}. Rewriting the entropy \eref{eq:sfinal} in the form
\begin{equation}
s(e,m) = \ln2+f(e,m)\,\arctanh\left[f(e,m)\right]-\case{1}{2}\ln\left[1-f(e,m)^2\right]
\end{equation}
and making use of \eref{eq:emvonhb1} and \eref{eq:emvonhb2}, we obtain as a final result for the canonical Gibbs free energy
\begin{equation}\label{eq:gbh}
-\beta g(\beta,h)=\beta e(\beta,h)-\case{1}{2}\ln\left[4-\frac{4}{\lambda_\perp}\left(\frac{h^2}{\lambda_\perp-\lambda_3}-2e(\beta,h)\right)\right],
\end{equation}
where $e(\beta,h)$ is a solution of \eref{eq:emvonhb2}. \Eref{eq:gbh} is valid for all $(\beta,h)$ satisfying the inequality \eref{eq:bhinequality}. For other values of inverse temperature and magnetic field, it is again the boundary \eref{eq:eofm} of the graph of $s(e,m)$ that determines the Legendre-Fenchel transform, and the result for $g(\beta,h)$ is the one derived in \sref{sec:nonconcave} and stated in \eref{eq:gfinal}. The curve
\begin{equation}
\beta h\lambda_\perp=(\lambda_\perp-\lambda_3)\,\arctanh\left(\frac{h}{\lambda_\perp-\lambda_3}\right)
\end{equation}
in the $(\beta,h)$-plane, at which the inequality \eref{eq:bhinequality} becomes sharp, determines the boundary separating the two different kinds of behaviour of $g$. This is precisely the line in the phase diagram separating the ferromagnetic phase of the Heisenberg magnet from the paramagnetic one, as plotted in \fref{fig:canPD} (right).

%------------------------------------
\section{Nonequivalence of ensembles}
\label{sec:nonequivalence}
%------------------------------------

In the previous section, we derived the canonical Gibbs free energy $g(\beta,h)$ from the microcanonical entropy $s(e,m)$. While, in the thermodynamic limit, it is always possible to compute canonical thermodynamic potentials from microcanonical ones, the inverse is not necessarily true. In those cases where the backward transition (from canonical to microcanonical) is also possible, one speaks of equivalence of ensembles, otherwise of nonequivalence. %Equivalence of ensembles is closely related to the geometric construction of planes touching the graph of a function which we used in \sref{sec:Legendre} to calculate the Legendre-Fenchel transform. If the set of planes $\bar{s}_{\beta,h,c(\beta,h)}(e,m)$, with $c(\beta,h)$ determined according to the description in \sref{sec:Legendre}, and for all $\beta$ and $h$, touches all points of $s$, equivalence of ensembles holds. If this is not the case, then there are certain values of $(e,m)$ which cannot be realized as canonical expectation values by fixing $\beta$ and $h$. This latter situation occurs precisely if and only if the microcanonical entropy is not a concave function of its arguments. See \cite{ElHaTur00,TouElTur04} for an introduction to the subject, and \cite{ElHaTur00} for a detailed account.

By inspection of rows three to seven in \fref{fig:entropy} [or by simple analysis of the results in \eref{eq:sfinal} and \eref{eq:domain}], the entropy $s$ for $\lambda_\perp>\lambda_3$ is seen to be a concave function on a domain which is a convex set. For $\lambda_\perp<\lambda_3$, the domain is not a convex set and therefore the entropy is neither convex nor concave. In the latter case, microcanonical and canonical ensembles are not equivalent, in the sense that it is impossible to obtain the microcanonical entropy $s(e,m)$ from the canonical Gibbs free energy $g(\beta,h)$ by means of a Legendre-Fenchel transform. This is evident from the fact that the outcome of a Legendre-Fenchel transform is always concave or convex \cite{Rockafellar}.

The physical interpretation of ensemble equivalence is that every thermodynamic equilibrium state of the system that can be probed by fixing certain values of $e$ and $m$ can also be probed by fixing the corresponding values of the inverse temperature $\beta(e,m)$ and the magnetic field $h(e,m)$. In the situation $\lambda_\perp<\lambda_3$, where nonequivalence holds, this is not the case: only equilibrium states corresponding to values of $(e,m)$ for which $s$ coincides with its concave envelope can be probed by fixing $(\beta,h)$; macrostates corresponding to other values of $(e,m)$, however, are not accessible as thermodynamic equilibrium states when controlling temperature and magnetic field in the canonical ensemble. In this sense, microcanonical thermodynamics can be considered not only as different from its canonical counterpart, but also as richer, allowing to probe equilibrium states of matter which are otherwise inaccessible. For more information on nonequivalent ensembles, see \cite{ElHaTur00} or the introductory article \cite{TouElTur04}.

There is a further long-range peculiarity, going under the name of partial equivalence \cite{CaKa07}, which can be observed in the anisotropic quantum Heisenberg model for any values of the coupling constants $\lambda_1$, $\lambda_2$, and $\lambda_3$. Partial equivalence here refers to the situation where a macrostate, associated with a certain pair of values $(e,m)$ in the microcanonical ensemble, corresponds to more than just one pair of values $(\beta,h)$ canonically. We have encountered this situation repeatedly when calculating the canonical Gibbs free energy $g$. In particular for the case $\lambda_\perp<\lambda_3$ in \sref{sec:nonconcave}, we computed ``touching planes'' $\bar{s}$ with different slopes $\beta$ and $-\beta h$, all touching the graph of the entropy $s$ at the same  boundary point $(-m_0^2\lambda_3/2,m_0)$. As a consequence, all pairs $(\beta,h)$ which solve equation \eref{eq:constraint} for the same value of $m_0$ correspond to the same microcanonical macrostate labelled by the parameters $(e(m_0),m_0)$. The same holds true for $\lambda_\perp>\lambda_3$ and values of $\beta$ and $h$ for which
\begin{equation}\label{eq:bhinequality2}
\beta \leqslant \frac{\lambda_\perp-\lambda_3}{h\lambda_\perp}\,\arctanh\left(\frac{h}{\lambda_\perp-\lambda_3}\right).
\end{equation}
This provides also an explanation for a peculiar observation reported in a study of the fidelity-metric of the Lipkin-Meshkov-Glick model, i.e.\ the special case $\lambda_3=0$ \cite{Scherer_etal09}. This article reports that, for $\beta$ and $h$ satisfying \eref{eq:bhinequality2}, a fidelity metric on the $(\beta,h)$-plane is not well defined. Having observed partial equivalence in this parameter regime, we know that entire curves, not single points, in the $(\beta,h)$-plane label one and the same macrostate. As a consequence, all (identical) macrostates along such a curve have distance zero from each other, which implies that a metric on the $(\beta,h)$-plane cannot be positive definite and is therefore not well defined.

%-----------------------------------------------------------------
\section{Thermodynamic equivalence of Heisenberg and Ising models}
\label{sec:thermoequiv}
%-----------------------------------------------------------------

It had been observed already in the 1970s that the Curie-Weiss isotropic Heisenberg model ($\lambda_1=\lambda_2=\lambda_3$) and the Curie-Weiss Ising model ($\lambda_1=0=\lambda_2$) are thermodynamically equivalent in the sense that their canonical Gibbs free energies coincide in the thermodynamic limit \cite{Niemeyer70}. From the results we have obtained in \sref{sec:Legendre}, it is obvious that an even more general statement can be made: For all anisotropy parameters satisfying $\lambda_\perp<\lambda_3$, the canonical Gibbs free energy $g(\beta,h)$ of the Curie-Weiss anisotropic Heisenberg model coincides with that of the Ising model. This is a direct consequence of the fact that, as pointed out in \sref{sec:nonconcave}, the Legendre-Fenchel transform of $s(e,m)$ for $\lambda_\perp<\lambda_3$ is determined exclusively by $s(-m^2\lambda_3/2,m)$, i.e.\ the entropy evaluated at the boundary $e(m)=-m^2\lambda_3/2$. Since the entropy at this boundary is identical to the entropy of the Curie-Weiss Ising model, thermodynamic equivalence holds in the canonical ensemble. Putting this differently, the microcanonical entropies $s(e,m)$ of the Curie-Weiss Ising model and the Curie-Weiss Heisenberg model with $\lambda_\perp<\lambda_3$ share identical concave envelopes, which implies that their Legendre-Fenchel transforms must be the same \cite{Rockafellar}. Remarkably, however, thermodynamic equivalence does not hold in the microcanonical ensemble, as is obvious from the different shapes of entropies in rows one to three of \fref{fig:entropy}.

Vertogen and de Vries have claimed in reference \cite{VertogenDeVries72} that the Curie-Weiss Ising model and the anisotropic Curie-Weiss quantum Heisenberg model are equivalent not only thermodynamically, but even on the level of their Hamiltonian operators in the thermodynamic limit. It is not obvious to the author how to reconcile this claim with the results of the present article: Equivalence on the operator level should imply identical thermodynamic functions also in the microcanonical ensemble, in conflict with the results of \sref{sec:micent}.

%----------------------------------------------------------------
\section{Physical relevance and experimental realization}
\label{sec:latticegas}
%----------------------------------------------------------------

In the thermodynamic limit, the microcanonical entropies \eref{eq:sNemdef} and \eref{eq:sem3def} discussed in this article describe the physical situation of fixed energy $e$ and fixed magnetization $m$. For neither of the two constraints it is immediately obvious how they can be realized in experiment: The quantum Heisenberg model was devised to model ferromagnetic spin systems which, in their traditional condensed matter realizations, are typically coupled to a thermal reservoir. As a consequence, the energy is not fixed, but fluctuates around a certain mean value, and the canonical ensemble is appropriate for a statistical equilibrium description of this situation.

Recently, however, it has been pointed out that cold atoms in optical lattices are an ideal laboratory for engineering systems which are governed by Hamiltonian operators formally equivalent to those of condensed matter spin systems. Furthermore, such cold atom realizations of condensed matter-type systems possess the very attractive feature of being highly controllable: by appropriately tuning Feshbach resonances and other parameters, the interaction type and strength can be tuned freely, even changing the character of the interaction force from attractive to repulsive. After switching off the cooling in such an experiment, total energy and number of atoms are conserved to a very good degree. As a consequence, a statistical description of such a lattice spin model should make use of the microcanonical ensemble.

%In a more direct way, long-range effects should be observable in long-range quantum spin systems undergoing a temperature-driven first-order transition. In this case, signatures of nonequivalent ensembles should show up in a microcanonical entropy $s(e)$, corresponding to the physical situation of conserved energy $e$, but freely fluctuating magnetization. In this situation, the above described detour via a lattice gas interpretation of the spin model would be dispensable.

The forces between cold atoms in optical lattices are typically of very short range: In most cases, s-wave scattering is dominant, and in this case it is appropriate to model the interaction force by an effective contact interaction. As a consequence, also the total magnetization is a conserved quantity, and the ensemble of constant energy and magnetization is appropriate for describing the equilibrium properties of such systems. A discussion of short-range spin systems under these conditions can be found in \cite{KaPlei09,Kastner09}.

Nonequivalence of ensembles is however restricted, as explained in the In\-tro\-duc\-tion, to long-range interacting systems. A long-range interaction can be introduced in cold gases by using atoms or molecules with a permanent electric or magnetic dipole moment, resulting in a dipole-dipole interaction potential decaying like $r^{-3}$ with the interparticle distance $r$. Alternatively, as shown by O'Dell {\em et al.}\ \cite{ODell_etal00}, an $r^{-1}$-interaction can be engineered by inducing, with appropriately tuned laser light, a dipole moment in atoms without a permanent dipole moment. Although such an algebraic decay is obviously different from the distance-independent forces in the Curie-Weiss-type model we have studied, it is known that Curie-Weiss-type models faithfully reproduce many properties of algebraically decaying long-range interactions qualitatively, and to some extent even quantitatively \cite{BisChayCraw06,Chayes09}. In particular, one can show under rather mild conditions that, by making the parameter $\alpha>0$ of an algebraically decaying $r^{-\alpha}$-interaction small, Curie-Weiss behaviour is approached continuously. Therefore, if the Curie-Weiss model shows a nonconcave entropy, the same is to be expected for algebraically decaying interactions with sufficiently small values of $\alpha$. Of course, this does not guarantee that nonequivalence of ensembles indeed persists to $\alpha=1$ or even $\alpha=3$, but it appears to be at least a plausible scenario.

It has been shown by Micheli {\em et al.}\ \cite{Micheli_etal06} that dipolar gases in optical traps allow for the realization of many Hamiltonians of interest in condensed matter physics, including the anisotropic quantum Heisenberg model. Experimentally, impressive progress in cooling and trapping cold dipolar atoms and molecules has recently been made \cite{Gries_etal05,Ni_etal08} and, although dipolar gases in optical lattices have not been realized at the time of writing, it seems just a matter of time until this goal will be achieved and the engineering of long-range interacting anisotropic quantum spin models will be possible. Unfortunately for our purposes, the dipole-dipole interactions allow for scattering transitions beyond s-wave scattering, and the total magnetization is therefore not a conserved quantity. As a consequence, the statistical ensemble realized in such an experimental setting is a microcanonical one with constant energy, but fluctuating magnetization. The entropy
\begin{equation}
s(e)=\max_m s(e,m)
\end{equation}
appropriate for describing this situation is, however, a concave function, and nonequivalence of ensembles cannot be detected under such conditions.

To probe nonequivalence in a cold atom experiment, we need to resort to a long-range spin model showing a temperature-driven {\em discontinuous}\/ phase transition: In this case, the microcanonical entropy $s(e)$ can be expected to be a nonconcave function, and an observation of this property would be possible under the experimental condition of conserved energy, but fluctuating magnetization. Such models will be investigated in a future work.

%----------------------------------------------------------------
\section{Conclusions}
\label{sec:conclusions}
%----------------------------------------------------------------

The main goal of the present article was to contribute towards the understanding of nonequivalence of ensembles in quantum spin systems. To this purpose, an exact, analytic calculation of the microcanonical entropy $s(e,m)$ of the anisotropic Curie-Weiss quantum Heisenberg model in the thermodynamic limit was reported. To the best of the author's knowledge, this is the first microcanonical calculation reported for a quantum spin system. With slight modifications, the strategy used to solve this problem, partly inspired by a canonical calculation by Tindemans and Capel \cite{TinCa74}, should also be applicable to other quantum systems with Curie-Weiss-type interactions, for example to Heisenberg models with spins larger than 1/2. From the microcanonical result, the canonical Gibbs free energy $g(\beta,h)$ was then recovered by means of a Legendre-Fenchel transform. This transform, especially in those instances when it does not coincide with a Legendre transform, also provides the key to understanding certain kinds of equivalences or nonequivalences that show up in the anisotropic Curie-Weiss quantum Heisenberg model: We found that the microcanonical entropy $s(e,m)$ is a nonconcave function for anisotropy parameters $\lambda_\perp<\lambda_3$, and in this case microcanonical and canonical ensembles are nonequivalent. Furthermore, independently of the values of the anisotropy parameters, partial equivalence occurs: different pairs $(\beta,h)$ and $(\beta',h')$ of the canonical variables can correspond to the same pair $(e,m)$ of microcanonical variables. Finally, the microcanonical result sheds light on the observation, dating from the 1960s and 1970s, that the Curie-Weiss Ising and the Curie-Weiss quantum Heisenberg models have identical canonical Gibbs free energies. We have shown that this thermodynamic equivalence also holds true in the anisotropic case for anisotropy parameters $\lambda_\perp<\lambda_3$, and it is a consequence of the peculiar nonconcave shape of $s(e,m)$ for these parameter values. Microcanonically, however, Heisenberg and Ising models are not thermodynamically equivalent, as is obvious from their differing microcanonical solution \eref{eq:sfinal}.

Microcanonical solutions of quantum spin model are argued to be relevant for a statistical description of dipolar gases in optical lattices. %The anisotropic quantum Heisenberg model is among the systems which can be engineered with cold polar molecules in optical lattices \cite{Micheli_etal06}, although with algebraically decaying long-range interactions instead of Curie-Weiss-type interactions discussed in the present article. 
In such experiments, the energy is controlled and conserved to a very high degree, rendering a mic\-ro\-can\-on\-i\-cal description adequate. However, the peculiarities of long-range systems, like nonconcave entropies, nonequivalence of statistical ensembles, or negative mic\-ro\-can\-on\-i\-cal response functions, should not depend on the precise nature of the long-range interactions. Still, a study of nonequivalent ensembles in quantum spin systems with algebraically decaying long-range interactions, as potentially realized in optical lattice experiments, is of course worthwhile and planned for future work.

These results and discussions point out the importance of nonstandard ther\-mo\-dy\-nam\-ics beyond the canonical ensemble for experiments with cold dipolar atoms or molecules in optical lattices: Equivalence of ensembles does not hold in general, and a comparison of experimental data with canonical statistical physical predictions is bound to fail in this case. On the other hand, the results show that such cold atom experiments can provide an ideal laboratory for studying fundamental issues of thermostatistics, like nonequivalence of ensembles, in a highly controlled environment.

\appendix

%----------------------------------------
%\section{Global maximum of $\mathcal{F}$}
%\label{sec:maxF}
%----------------------------------------

%-------------------------------------------------------------
\section{Evaluation of $\mathcal{F}$ at the stationary points}
\label{sec:evalF}
%-------------------------------------------------------------
It is shown how to evaluate $\mathcal{F}$ as given in \eref{eq:F2} at a stationary point determined by equations \eref{eq:saddle3a}--\eref{eq:saddle3d}. There are several ways to satisfy \eref{eq:saddle3d} for $\alpha=1,2$:
\begin{enumerate}
\item $x_1=0$ and $ms(\lambda_3-\lambda_2)=t$,
\item $x_2=0$ and $ms(\lambda_3-\lambda_1)=t$,
\item $ms(\lambda_3-\lambda_1)=t$ and $ms(\lambda_3-\lambda_2)=t$,
\item $x_1=0$ and $x_2=0$.
\end{enumerate}
We discuss (i) in detail and argue later that the other cases do not contribute anything new.

Assuming that $x_1=0$ and $ms(\lambda_3-\lambda_2)=t$, and making use of \eref{eq:saddle3c}, equation \eref{eq:Rt1} simplifies to
\begin{equation}\label{eq:Rt2}
R_t=\sqrt{\lambda_2\left( x_2^2 +m^2s^2\lambda_2\right)},
\end{equation}
and the set of equations \eref{eq:saddle3a}--\eref{eq:saddle3d} takes on the form
%\numparts
\begin{eqnarray}
0&= 2es^2+x_2^2+x_3^2 = s^2\left(2e+m^2\lambda_3\right)+x_2^2,\label{eq:saddle4a}\\
0&= m\left(R_t-s\lambda_2\tanh R_t\right).\label{eq:saddle4b}
\end{eqnarray}
%\endnumparts
From \eref{eq:saddle4a} it follows that solutions exist only under the condition
\begin{equation}
2e+m^2\lambda_3<0.
\end{equation}
Inserting \eref{eq:saddle4a} into \eref{eq:Rt2} we obtain
\begin{equation}\label{eq:Rt3}
\frac{R_t}{s\lambda_2}=\sqrt{m^2\left(1-\frac{\lambda_3}{\lambda_2}\right)-\frac{2e}{\lambda_2}},
\end{equation}
which allows us to write $\mathcal{F}$, given in \eref{eq:F2}, in the form
\begin{equation}\label{eq:F3}
\mathcal{F}=es+mt-\underbrace{\frac{x_2^2+x_3^2}{2s}}_{\displaystyle=-es}-\frac{1}{2}\ln\left[1-\left(\frac{R_t}{s\lambda_2}\right)^2\right].
\end{equation}
Using $mt=m^2s(\lambda_3-\lambda_2)$ and \eref{eq:saddle4b}, we can rewrite the first three terms on the right hand side of this equation as
\begin{equation}
\fl s\left[2e+m^2\left(\lambda_3-\lambda_2\right)\right] = -\frac{R_t^2}{s\lambda_2} = -\frac{R_t}{s\lambda_2}\arctanh\left(\frac{R_t}{s\lambda_2}\right) = \frac{R_t}{2s\lambda_2}\ln\left(\frac{1-\frac{R_t}{s\lambda_2}}{1+\frac{R_t}{s\lambda_2}}\right).
\end{equation}
Inserting this expression into \eref{eq:F3} and making use of \eref{eq:Rt3}, we obtain as a final result \eref{eq:Ffinal} and \eref{eq:fem}, where $\lambda_\perp\equiv\lambda_2$. Real solutions for $\mathcal{F}$ exist only when the argument of the logarithm is positive,
\begin{equation}
1>\left(\frac{R_t}{s\lambda_2}\right)^2 = m^2\left(1-\frac{\lambda_3}{\lambda_2}\right)-\frac{2e}{\lambda_2},
\end{equation}
which leads to a second inequality to be satisfied by $e$ and $m$.

Case (ii) of the above list yields the same result, but with the roles of $\lambda_1$ and $\lambda_2$ interchanged, i.e.\ $\lambda_\perp\equiv\lambda_1$. Case (iii) can only be satisfied if $\lambda_1=\lambda_2$, but the result for $\mathcal{F}$ is the same as in (i) or (ii). Case (iv) finally has solutions only for $2e=-m^2\lambda_3$, but the corresponding results for $\mathcal{F}$ are again the same as in the cases (i) and (ii).

%-------------------------
\vspace{3mm}
\bibliographystyle{unsrt}
\bibliography{CWHeisenberg}

\end{document}